\documentclass[12pt]{article}
\pdfoutput=1
\usepackage[top=1in]{geometry}
\usepackage{latexsym}
\usepackage{amsmath}
\usepackage{amsfonts}
\usepackage{amssymb}
\usepackage{graphicx}
\usepackage{multirow}
\usepackage{color}
\usepackage{tikz}
\usepackage{pdflscape}
\usepgflibrary{arrows}

\title{Chiral properties of graphene h-BN hybrid systems}
\author{A. Molenda$^1$,I. Zasada$^1$\thanks {e-mail:izasada@wfis.uni.lodz.pl}, P. Ma\'slanka$^2$\\
$^1$\small Department of Solid State Physics,
 Faculty of Physics and Applied Informatics\\
$^2$\small Department of  Computer Science, 
 Faculty of Physics and Applied Informatics\\
\small University of \L\'od\'z,\\
\small Pomorska 149/153, 90-236 {\L}\'od\'z, Poland}

\date{}
\begin{document}
\maketitle
\begin{abstract}
The application of the chiral decomposition procedure to hybrid graphene h-BN systems revealed rules for the partition of the system into effective subsystems being bilayers plus monolayer in case the number of layers is odd. Three types of subsystems have been detected namely purely graphene bilayers and monolayers, mixed bilayers and pure h-BN monolayers depending on the hybrid composition. The effective parameters characterizing these chiral subsystems consist of the interlayer couplings and on-site potentials which shows the mechanism of compensation of the asymmetry introduced into the system by h-BN layers. For illustration, we provide a pedagogical overview about chiral tunneling in graphene subsystems (MLG, BLG) present in hybrid with one h-BN layer. We have established the parameter ranges for which the characteristic features in the spectrum are observed, such as Fabry-P\'erot resonances in the case of MLG and "magic angles" in the case of effective BLG. We also consider different hybrid stacking in order to indicate effective systems with the desired properties required in the electronic and spintronic applications. 

\end{abstract}

\newpage
\section{ Introduction}

Graphene has attracted intense experimental and theoretical efforts since its first isolation and identification in 2004 \cite{b1}. Experimental studies have revealed its exotic transport properties, such as perfect transmission for electrons incident in the normal direction at a potential barrier (Klein paradox) \cite{b2} or an anomalous quantum Hall effect \cite{b3}. All these results can be in general explained by the two-dimensional massless Dirac equation \cite{b4} $\div$ \cite{b7}. Multilayer graphene which consists of few stacked graphene layers also attract attention driven by advances in material preparation \cite{b8} $\div$ \cite{b10}  as well as by the unusual electronic properties of these systems \cite{b11} $\div$ \cite{b18}. A variety of stacking structures in multilayer systems coming from the production method gives different electronic properties. For example, micromechanical cleavage of graphite leads to the Bernal stacking type which provides a mixture of effective bilayers plus monolayer if the number of layers in the system is odd \cite{b14, b19}. The properties of the effective bilayer subsystems are connected with chirality of charge carriers in each specific multilayer system. This result implies that it may be possible to find different chiral fermions in the multilayer graphene systems combined with other 2D materials.\\ 
Since 2010, hexagonal boron nitride (h-BN) supported graphene samples have activated a new rush in graphene research \cite{b20} $\div$ \cite{b24}. Hexagonal boron nitride is a layered material similar to graphite, which can also be exfoliated to give samples with atomically flat surfaces. Two triangular sublattices consist in this case of boron and nitrogen atoms, respectively. The lattice mismatch between h-BN and graphene is $ \sim 1,8 \% $ which can give the uniform heterostructures \cite{b25}. In contrast to graphene, in h-BN the boron and nitrogen atoms form ionic bonds which results in a large band gap of $\sim 6 $ eV \cite{b26}. Thus, the use of h-BN in modeling the properties of graphene multilayers guarantees that the conducting channels would be only through graphene bands. Moreover, transport measurements of graphene on h-BN devices show an improvement of charge carrier mobility by a factor of three to ten compared to the graphene on $\text{SiO}_2$ devices \cite{b20}.\\ 
Very recently, we presented \cite{b27} the generalization of the chiral decomposition procedure \cite{b19} for multilayer graphene supported by h-BN layer. We found that the $N$-layer graphene Bernal stacking system deposited on h-BN layer can be described by isolated $N/2$  effective bilayer systems with one effective h-BN layer if $N$ is even or $(N-1)/2$   effective bilayers plus one MLG modified by h-BN layer if $N$ is odd. The electronic properties of the effective bilayer systems depend strongly on their local surrounding so they can be tuned by the appropriate change of layer arrangement within the multilayer heterostructures. On the other hand, there is a generic interest in the possibilities of engineering the specific properties of graphene systems in view of application in nanoelectronics. In particular, bilayer graphene (BLG) was proposed as a non-magnetic, pseudospin-based version of a spin valve, in which the pseudospin polarization in neighboring regions of a graphene bilayer is controlled by external gates \cite{b28}. Bilayer graphene possesses a pseudospin degree of freedom, which is associated with the electron density difference between the top layer and the bottom layer. If the layer degree of freedom is considered as a pseudospin, an external potential difference between the two layers creates a pseudospin polarization, which corresponds to the charge transfer between the two layers and the polarization direction is along the normal to the graphene plane. It seems promising to develop bilayer graphene "pseudospintronics" by utilizing the layer pseudospin \cite{b29}. In this context, the BLGs with specific electronic properties can be required. On the other hand, the h-BN support does not allow to keep the specific properties of  monolayer graphene. MLG present in the multilayer with odd number of graphene sheets always exhibits the small band gap which can be undesirable for electronic applications.\\ 
In the present paper, we consider how the differences in placement of the h-BN layer within the multilayer graphene stacking system appear in the physical properties of the effective subsystems. By using the chiral decomposition procedure proposed in Ref. \cite{b27}  we could reveal some general rules in arrangement of these heterostructures which lead to the same qualitative results. We discuss then the quantitative differences within specific group. For illustration we consider the tunneling phenomena. We restrict to the electron transport through the potential barrier higher than the incident electrons energy in order to observe interband tunneling when electrons outside the barrier (conduction band) transforms into holes inside it (valence band), or vice-versa \cite{b2}. Moreover, we discuss the properties of graphene h-BN hybrid systems supported by h-BN.\\
The rest of the paper is organized as follows. Section II gives the details of computational methods. Results are discussed in Section III and a summery is given in Section IV.

\section{Model considerations}

Graphene is formed in hexagonal symmetry with two triangular sublattices which leads to the band structure holding two gapless points. The low energy properties of this system are described  by the 2D massless Dirac equation. For the multilayer graphene, one consider network of the Dirac fermions systems interacting via interlayer hopping parameters.   According to Density Functional Theory (DFT) \cite{b30}, \cite{b31} and experimental findings \cite{b32}, it is energetically favorable for the atoms of sublattice A(B) to be displaced along the honeycomb edges in a way that an atom from the sublattice A(B) sits on top of an atom belonging to another sublattice B(A). This stacking rule implies the three distinct but equivalent projections of the 3D layered structures onto x-y plane and   distinct N-layer stack sequences \cite{b33}. It is also known that there are mainly two stacking types for graphite: the so-called Bernal stacking, forming the layer sequence $1212...$. and the rhombohedral stacking which form the layer sequence $123123...$. However, since the rhombohedral systems are considered to be unstable against external perturbations \cite{b14}, calculations in this paper is devoted to Bernal stacking systems. All the more, this staking order is also favorable for multilayer h-BN system \cite{b34, b35}. The multilayer systems can be described by the tight-binding π bands Hamiltonian which for the Bernal stacking of arbitrary number of graphene and/or h-BN monolayers is given by:

\begin{equation}
\label{e1}
H_N=\left(  
\begin{smallmatrix} 
\varepsilon_{1,\alpha}&-\gamma_{1,0}f_1(\vec{k})& 0 &\gamma_{1,2}&0&0&0&0&\\
-\gamma_{1,0}f_1^*(\vec{k})&\varepsilon_{1,\beta} &0& 0&0&0&0&0&\\
   0&0&\varepsilon_{2,\alpha}&-\gamma_{2,0}f_2(\vec{k})& 0 &0&0&0&\\
\gamma_{1,2}& 0 & -\gamma_{2,0}f_i^*(\vec{k})&\varepsilon_{2,\beta}&\gamma_{2,3}&0&0&0&\\
0&0&0&\gamma_{2,3}&\varepsilon_{3,\alpha}&-\gamma_{3,0}f_3(\vec{k})&0&\gamma_{3,4}\\
0&0&0&0&-\gamma_{3,0}f_i^*(\vec{k})&\varepsilon_{3,\beta}&0&0\\
0&0 & 0& 0& 0& 0&\cdots\\
0&0&0&0&\gamma_{3,4}& 0&&\cdots\\
&&&&&&&&\ddots        
\end{smallmatrix}
\right) 
\end{equation}

The diagonal terms $\varepsilon_{i,\alpha(\beta)}$\  denote the on-site energy of electron at the atom in layer $i$ belonging to sublattice A(B). In the first approximation, they can be equal to the energy of an electron in the $2p_z$\ orbital of an atom. However, this energy is modified as atoms bond together forming the lattice and can be considered as a parameter to fit with the experimental findings. Two parameters $\gamma$\  describe the strength of the coupling between a specific pair of atoms: $\gamma_{i,0}$\  denotes the coupling between the nearest neighbors in each monolayer while $\gamma_{i,i+1}=\gamma_{i+1,i}$\  describes direct interlayer coupling. Lets notice that next-nearest neighbor couplings are non-essential for the problem at hand, and are therefore neglected for simplicity. The geometrical factor $f(\vec{k})$\   resulting from a summation over nearest neighbors in each monolayer can be written in term of vectors $d_i$\  describing geometric relation between two sublattices and has the following form:

\begin{equation}
\label{e2}
\begin{split}
f(\vec{k})\equiv\sum_{i=1}^3exp(i\vec{k}\cdot\vec{d_i})=exp\left(i\frac{•k_ya}{\sqrt{3}}\right) +2exp\left(-i\frac{•k_ya}{2\sqrt{3}}\right)cos\left( \frac{k_xa}{2}\right) 
\end{split}
\end{equation}

It can be approximated by:$f(\vec{k})\approx-\frac{\sqrt{3}a}{2\hbar}(\xi p_x-ip_y)$ with  $\vec{p}=\hbar \vec{k}-\hbar\vec{K_\xi}, \xi=\pm1$. In the approximation based on nearest neighbors interactions, the overlap between two nearest neighbor atoms in monolayer and the overlap between the atoms which are directly above/below each other in neighboring monolayers should be taken into account. However, in all situations considered in this paper, the overlaps are neglected due to their small values. Thus, the band energies may be determine from the eigenvalue equation by solving the secular equation det $(H-EI)=0$ . In the present case, the recurrence relation for determinant of $N$-layer system \cite{b27} can be written in the following form:

\begin{equation}
\label{e3}
\begin{split}
&det(H_N-EI)=D_N=(\varepsilon_{N,\alpha}\varepsilon_{N,\beta}-\gamma^2_{N,0}\vert f_N(\vec{k})\vert^2) \cdot D_{N-1} \\
&+(\varepsilon_{N,\beta}-E)\left\lbrace \sum_{k=1}^{N-2}(-1)^k(\varepsilon_{N-k,\alpha}-E)\prod_{i=N-k}^{N-1} (1-s_i)^2\gamma_i^2D_{N-k-1}\right.\\
&\left.+(-1)^{N-1}(\varepsilon_{1,\alpha}-E)\prod_{i=1}^{N-1}(1-s_i)^2\gamma_i^2\right\rbrace \\
&+(\varepsilon_{N,\alpha}-E) \left\lbrace \sum_{k=1}^{N-2}(-1)^k(\varepsilon_{N-k,\beta}-E)\prod_{i=N-k}^{N-1}s_i^2\gamma_i^2D_{N-k-1}\right.\\
&\left.+(-1)^{N-1}(\varepsilon_{1,\beta}-E)\prod_{i=1}^{N-1}s_i^2\gamma_i^2 \right\rbrace 
\end{split}  
\end{equation}
where

\begin{align}
\label{e4}
s_i&=
\left\{
\begin{array}{ll}
0  &\text{if} \ \ i\text{ is even number} \\
1  &\text{if} \ \ i\text{ is odd number}
\end{array} \right. 
\end{align}

In order to calculate physical quantities of multilayer systems, it is convenient to factorize the original Hamiltonian with $2N\times2N$  matrix form into an effective Hamiltonian which is composed of block diagonalized elements. The effective Hamiltonian is derived by assuming det$(H-EI)=\text{det}(H^{eff}-EI)$. This is exact mapping of original Hamiltonian without using any approximation. In the case of multilayer graphene, $N$-layer Bernal stacking system can be described by $N/2$ bilayer systems with effective interlayer hoppings and one monolayer system if $N$\ is odd \cite{b14, b19} while the $N$-layer graphene Bernal stacking system deposited on h-BN layer can be described by isolated $N/2$ bilayer systems with some effective interlayer hopping and onsite energies and one h-BN with effective onsite energy if $N$ is even or $(N-1)/2$  bilayer plus one MLG modified by h-BN layer if $N$ is odd. In these last cases the MLG/h-BN bilayer is characterized by the effective interlayer hopping and effective onsite energy in h-BN sublattice which interact with MLG sublattice \cite{b27}. We use the same decomposition procedure here to analyze graphene h-BN heterostructures in two cases:

\begin{itemize}
\item[(i)]  with one h-BN layer placed in the arbitrary position within multilayer graphene system, and
\item[(ii)] with two or more h-BN layers, one distributed inside the graphene system and the others playing role of the support. 
\end{itemize}

At a first stage of considerations we calculate the band structures. The values of all parameters employed in the present numerical calculations are taken from the standard set of parameters used in the theoretical analysis of the graphene/h-BN multilayer systems properties \cite{b23, b34}. Within the error bar they agree with the experimentally estimated parameters \cite{b36}. Two parameters describing the strength of the coupling between a specific pair of atoms are taken to be: $\gamma_{i,0\mathrm{(C-C)}}\equiv\gamma_0=3.033$ eV  and $\gamma_{i,0\mathrm{(C-N)}}\equiv\gamma'_0 =0.25$ eV for coupling between the nearest neighbors in each graphene and h-BN monolayer, respectively, while $\gamma_{i,i+1 \mathrm{(C-C)}} =0.39$ eV  and $\gamma_{i,i+1 \mathrm{(C-N)}} =0.25$ eV for interlayer coupling. Note, that we consider only stockings with Carbon-Nitrogen interaction \cite{b34}. Moreover, it is assumed here that lattice constant   is common for all monolayers building a system. Thus, the difference between lattice constant of graphene and h-BN is not taken into account here. However, it seems to be justified by the fact that we consider the ultrathin multilayer systems which are willing to accommodate \cite{b25}  in contrast to graphene on h-BN substrate when moir\'e  pattern is observed \cite{b26}. Furthermore, we take the onsite energy to be equal zero for Carbon, $3.36$ eV for  Nitrogen and $-3.66$ eV for Boron \cite{b34}. 
Next, for illustration how the differences of stacking structures appear in the physical quantities we deal with transport properties. We consider the charge carriers incident on a potential barrier. Particles traveling from the left to the right in multilayer system are incident at energy  $E>0$ on a potential barrier of height $U$  and width $d$. The plane-wave solution of the Schr\"odinger equation $H\Psi=(E-U)\Psi$  describes this situation \cite{b27}.

\section{Results and discussion }

As described in Section 2, we assume the nearest neighbors interactions and not locally modified onsite energies. It means we neglect the asymmetry leading to "maxican hat" in band structure, the trigonal warping and opening up of small energy gaps between the conduction and the valence bands. The details of the band structure including all these effects is a very complicated problem that has yet to be studied. In view of the large uncertainty in the parameters involved, it is meaningless to introduce, at this stage, a more complicated model. However, it should be underline that inclusion of the other parameters does not cause principal problems, but the analysis becomes more complicated. For example, the inclusion of trigonal warping phenomenon does not block the possibility of chiral decomposition of the Hamiltonian describing the graphene multilayer systems with arbitrary number of layers \cite{b19}. Hence, we believe that the Hamiltonian (1) correctly captures the main features of the present problem.
 
\subsection{h-BN layer within graphene multilayers}

\begin{figure}[htbp]
\includegraphics[clip, trim=2cm 21.5cm 2cm 2cm,scale=0.9]{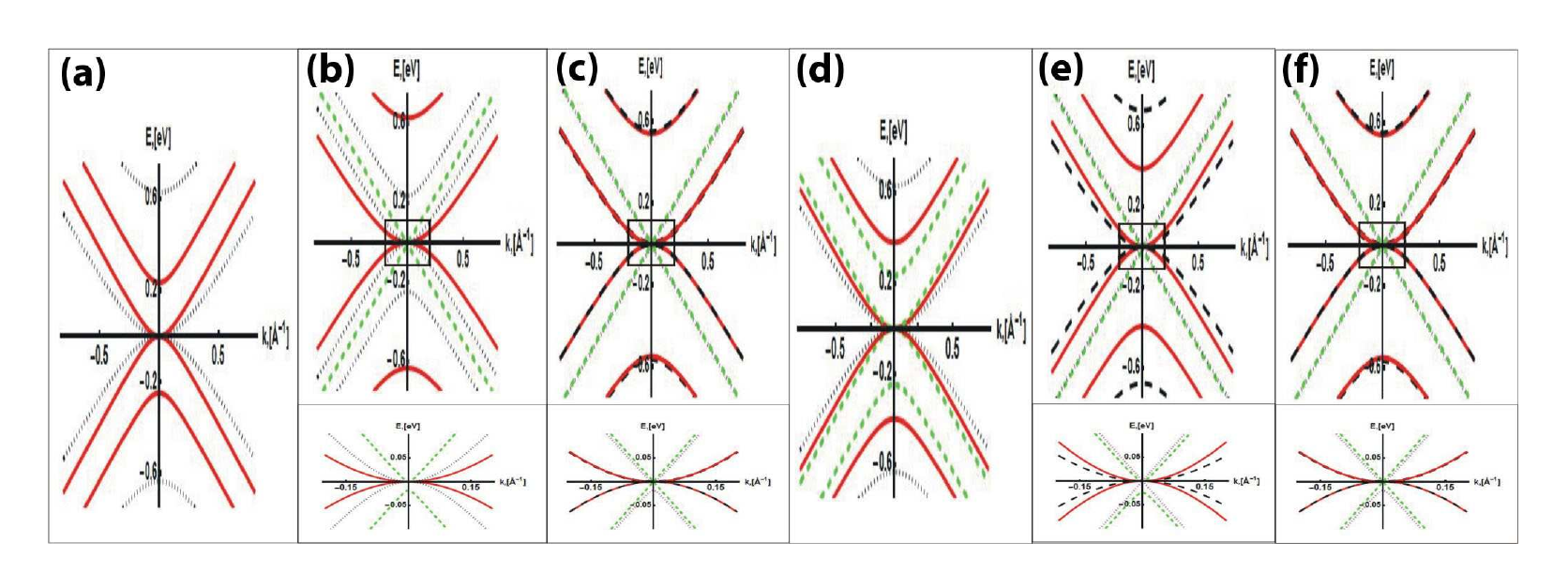}
\caption{\label{fig1}(Color online)  Band structure of graphene h-BN hybrid systems: a) 4GL/h-BN; b) 4GL/h-BN/MLG; c) TLG/h-BN/BLG; d) 4GL/h-BN/BLG; e) 5GL/h-BN/MLG; f) TLG/h-BN/TLG. For clarity, only the graphene bands are shown. Colors indicate the effective subsystems existing in individual hybrids. Insets present the band structure in the vicinity of K point as indicated by the black windows.}
\end{figure}

The h-BN layer can be placed inside multilayer graphene system with a total number of graphene layers $N=N_1+N_2$, in three ways: $N_1$-even/h-BN/$N_2$-even; $N_1$-odd/h-BN/$N_2$-odd; $N_1$-even/h-BN/$N_2$-odd. In order to investigate changes in the band structure induced by different positions of the h-BN layer within graphene/h-BN system, let us start with four graphene layers deposited on h-BN and create a new systems by adding graphene layers. The effective Hamiltonian for 4LG/h-BN is given in Appendix A by eq.(\ref{e21})
and its band structure is presented in Fig.\ref{fig1}a. Two effective BLG can be recognized. By adding next MLG we go to the case of $N_1$-even/h-BN/$N_2$-odd system type. We consider two different arrangements: 4LG/h-BN/MLG and TLG/h-BN/BLG. These two stacking systems are described by the  Hamiltonians (\ref{e22}), (\ref{e23}) and their band structures are shown in Fig.\ref{fig1}b and \ref{fig1}c, respectively.
Additional graphene layer will create two other types of systems, namely: $N_1$-even/h-BN/$N_2$-even and $N_1$-odd/h-BN/$N_2$-odd. The effective Hamiltonians for 4LG/h-BN/BLG, 5LG/h-BN/MLG and TLG/h-BN/TLG can be written in the form (\ref{e24}), (\ref{e25}) and (\ref{e26}), respectively. The band structures of these systems are shown in Fig.\ref{fig1}d, \ref{fig1}e and \ref{fig1}f, respectively. 

\begin{table}
\caption{\label{tab1} Effective parameters $(\gamma_{\text{\it eff}},\varepsilon_{\text{\it eff}})$ of the effective subsystem Hamiltonian of the Bernal stacking hybrid systems. Energy gap $\Delta_g$ induced by h-BN.}
\begin{center}
\begin{tabular}{|c|c|r|r|r|}
\hline
Original multilayer & After decoupling & 
\multicolumn{1}{|c|}{$\gamma_{\text{\it eff}}$ [eV]} &
\multicolumn{1}{|c|}{$\varepsilon_{\text{\it eff}}$ [eV]} &
\multicolumn{1}{|c|}{$\Delta_g$ [eV]} \\
\hline\hline
\multirow{3}{*}{5GL/h-BN} &
 BLG & 0.675 & -0.0016 &\\	
 \cline{2-5}
& BLG & 0.389& -0.0047&\\
\cline{2-5}	
&MLG/h-BN&0.145&3.3704&0.0062\\
\hline
\multirow{4}{*}{5GL/h-BN/MLG} &
BLG&0.675&-0.0016&{}\\
\cline{2-5}	
&BLG	&0.389&-0.0047&\\
\cline{2-5}
&MLG	&&&\\
\cline{2-5}
&MLG/h-BN&0.288&3.3700&0.0246\\
\hline
\multirow{5}{*}{5GL/h-BN/TLG} 
&BLG&0.675&-0.0016&\\
\cline{2-5}	
&BLG&0.389&-0.0047&\\
\cline{2-5}	
&BLG&0.551&-0.0047&\\
\cline{2-5}	
&MLG&&&\\
\cline{2-5}
&MLG/h-BN&0.229&3.3798&0.0154\\
\hline
\multirow{6}{*}{5GL/h-BN/5LG} 
&BLG&0.677&-0.0014&\\
\cline{2-5}	
&BLG&0.673&-0.0018&\\
\cline{2-5}	
&BLG&0.394&-0.0043&\\
\cline{2-5}	
&BLG&0.385&-0.0050&\\
\cline{2-5}
&MLG&&&\\
\cline{2-5}
&MLG/h-BN&0.204&3.3828&0.0123\\
\hline
\multirow{2}{*}{TGL/h-BN}
&BLG	&0.551&-0.0047&\\
\cline{2-5}	
&MLG/h-BN&0.177&3.3670&0.0093\\
\hline
\multirow{4}{*}{TGL/h-BN/TLG}
&BLG&0.556&-0.0041&\\
\cline{2-5}	
&BLG&0.546&-0.0053&\\
\cline{2-5}	
&MLG&&&\\
\cline{2-5}
&MLG/h-BN&0.250&3.3760&0.0185\\
\hline
\end{tabular}
\end{center}
\end{table}

The simple inspection of Fig.\ref{fig1} allows to reveal the following general rules:
\begin{itemize}

\item[1.] $N_1$-even/h-BN/$N_2$-odd ($N_1$-odd/h-BN/$N_2$-even) type of system decouples in $(N-1)/2$ BLG plus MLG/h-BN. It is qualitatively the same as in the case of odd number of graphene layers supported by one h-BN layer \cite{b27}. 

\item[2.] $N_1$-even/h-BN/$N_2$-even type of system decouples in  $N/2$  BLG. It is qualitatively the same as in the case of even number of graphene layers supported by one h-BN layer \cite{b27}.

\item[3.] $N_1$-odd/h-BN/$N_2$-odd type of system decouples in $(N-2)/2$  BLG plus one MLG plus MLG/h-BN. It represents new situation in which the unusual properties of MLG are preserved.
\end{itemize}

\begin{figure}
\includegraphics[clip, trim=2cm 19.5cm 2cm 2cm,scale=0.9]{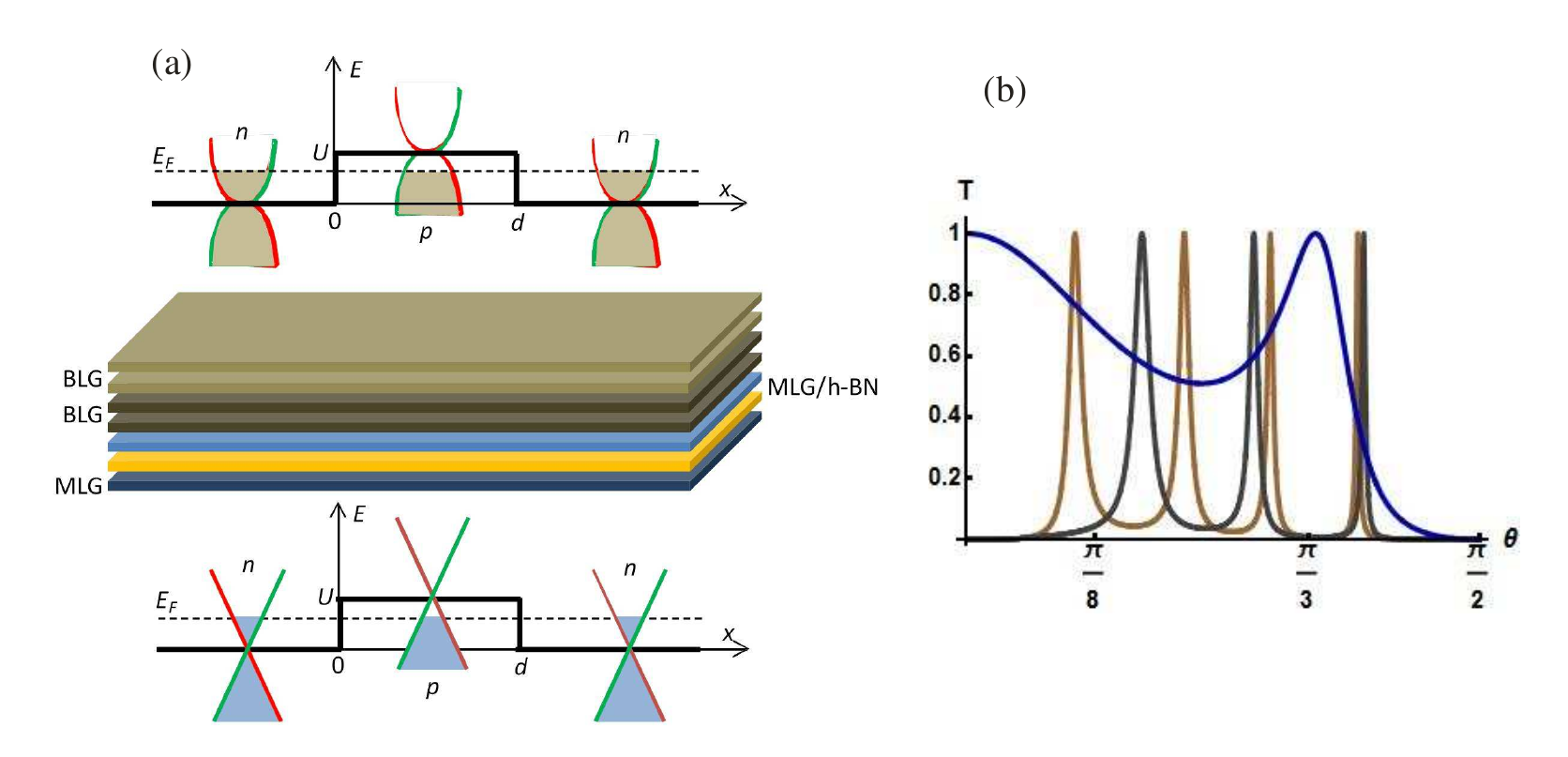}
\caption{\label{fig2}(Color online)  (a) Schematic view of the band structures across a sharp npn junction for MLG and effective BLG. (b)Transmission probabilities as a function of incidence angle for three effective subsystems present in 5GL/h-BN/MLG hybrid. There are three modes: two for effective BLG's (brawn and light brown curves) and one for MLG (blue curve).}
\end{figure}

The effective parameters for few examples of the systems from the last case are gathered in Table I. For better analysis, the effective parameters characterizing 5LG/h-BN and TLG/h-BN are also put in the Table.\\
First of all, let us remark that the low-energy electronic structure of graphene h-BN hybrid systems consists of chiral pseudospin doublets similar to graphene multilayers \cite{b33}. The rule of partitioning the system is based on the following steps: i) identification of segments within which there are no reversals of stacking sense (all possible graphene bilayers in the present case of Bernal type arrangement); ii) partition of the remaining segments into smaller elements excluding layers contained within previously identified partitions (MLG/h-BN bilayer and/or one h-BN(graphene) monolayer in the present case). Next, let us underline that the effective on-site potentials $\varepsilon_{eff}$\  induced in graphene bilayers act as compensation against the imposed asymmetry by h-BN layer and do not give rise to the gap in the spectrum (see Fig.\ref{fig1}). The energy band gaps appear in the graphene monolayers coupled with h-BN and their values depend on the multilayer composition. It is interesting that the effective subsystems in G/h-BN/G hybrids mimic their properties from G/h-BN systems. For example, two BLG subsystems from 5GL/h-BN/TLG are characterized by the same effective parameters as BLGs in the 5GL/h-BN while the third BLG can be found in TGL/h-BN. However, a crucial difference is that one of the graphene multilayer keeps its original properties as if the rest of system did not exist.
For all other systems considered here similar regularities can be easily recognized. Moreover, multilayers symmetric with respect to h-BN layer position are decoupled into symmetric BLG subsystems (see Fig.\ref{fig1}f and Table \ref{tab1}). Thus, one can have the pairs of almost identical effective bilayers in one stack.\\
For illustration, we discuss transport properties through a square potential barrier for the effective subsystems present in 5GL/h-BN/MLG (see Fig.\ref{fig1}e and Table \ref{tab1}). Schematic view of the band structures across a sharp npn junction for MLG and effective BLG is presented in Fig.\ref{fig2}a while in Fig.\ref{fig2}b we can see the transmission probabilities as a function of incidence angle for all subsystems calculated in the case of the following barrier parameters: width 100 nm; height 50 meV and the energy of the incidence electron equal 17 meV. It can be noticed that it is not easy to distinguish between transmission modes  from different subsystems. In particular, transmission for BLG's is always mixed up with transmission for MLG mod. On the other hand there is a quite wide range of incidence angles for which only MLG mod is observed.  
 
\begin{figure}
\includegraphics[clip, trim=2cm 16cm 2cm 2cm,scale=0.9]{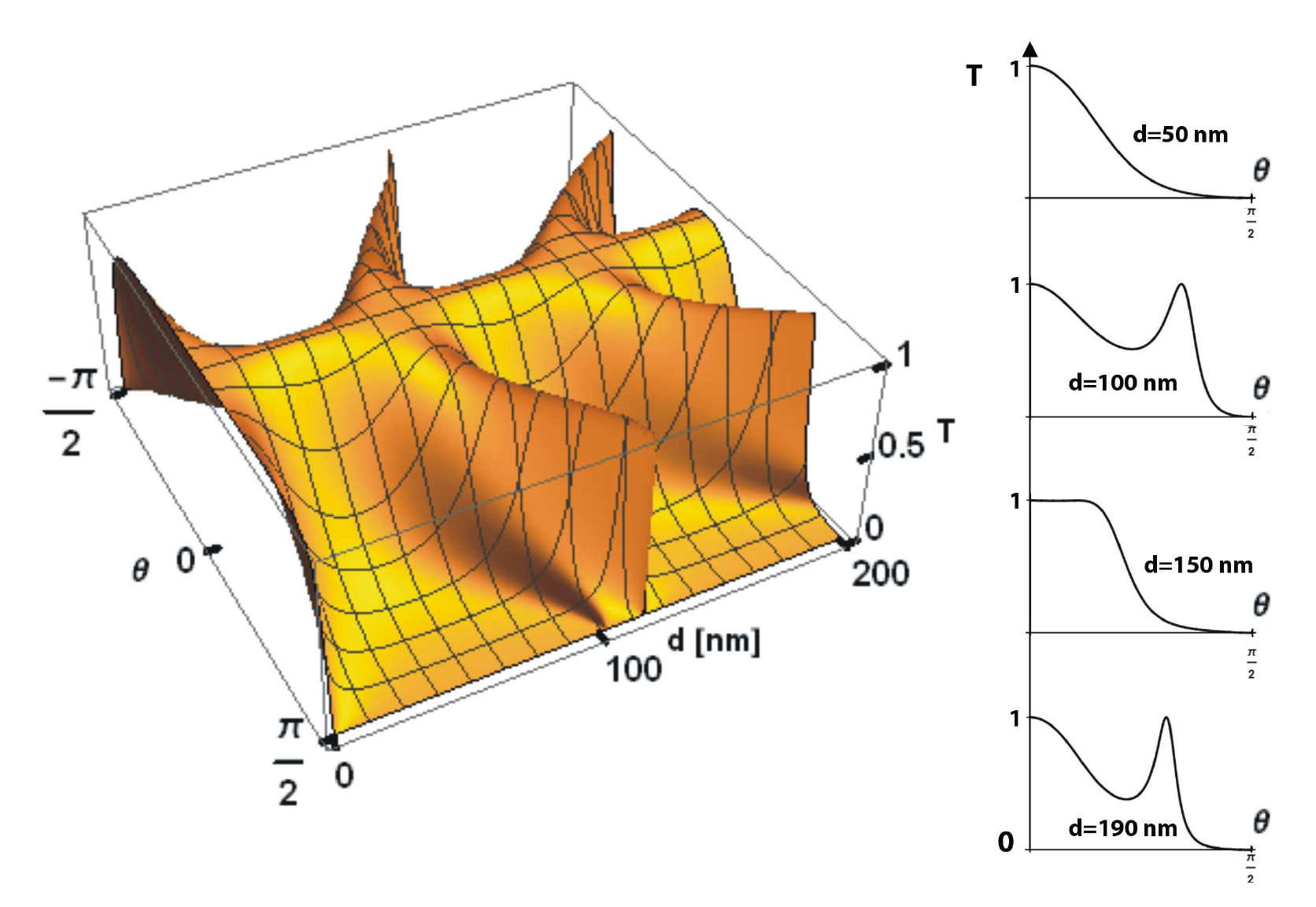}
\caption{\label{fig4}(Color online)  Transmission probability as a function of the incidence angle $\theta$ and the barrier width for a fixed energy of incidence electrons ($E=$ 17meV) below the barrier height ($U =$ 50meV) calculated for MLG existing in 5GL/h-BN/MLG system (gray dotted lines in Fig.\ref{fig1}e). Right panel shows the cross-sections for four characteristic barrier widths for which the Fabry-P\'erot resonances are present and absent, respectively.}
\end{figure}

\begin{figure}
\includegraphics[clip, trim=2cm 15.5cm 2cm 2cm,scale=0.9]{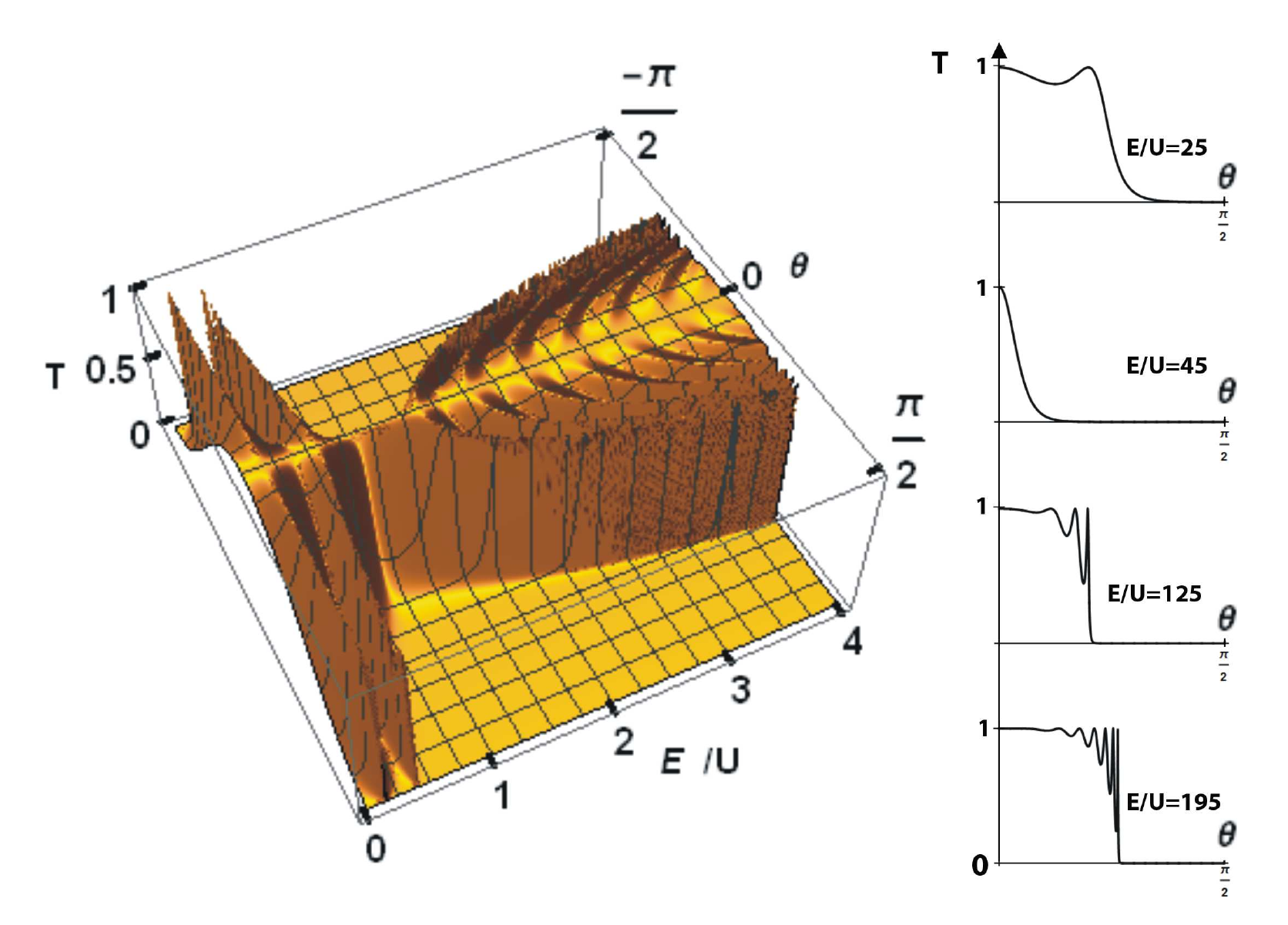}
\caption{\label{fig5}(Color online)  Transmission probability as a function of the incidence angle θ and the dimensionless energy $E /U$ ($U =$ 50meV) for a fixed width of the barrier ($d =$ 100nm) calculated for MLG existing in 5GL/h-BN/MLG system (gray dotted lines in Fig.\ref{fig1}e). Right panel shows the cross-sections for four characteristic regions. }
\end{figure}

Let us start the detailed analysis with graphene monolayer.  
In Fig.\ref{fig4} we show the dependence of transmission coefficient on electrons incidence angle and barrier width for the fixed energy of the incidence electrons below the barrier height ($0<E<U$).
This corresponds to npn junction in graphene. We can recognized two regions: one with petal-like shape of $T(\theta)$ when plotted as a function of the incidence angle for a fixed barrier width from certain ranges of values (see second and fourth plot in right panel), and second with smooth shape of $T(\theta)$ for the barrier width from other ranges of values (see first and third plot in right panel). In the first case we have to do with Fabry-P\'erot resonances $T(\theta\neq0)=\ 1$ \cite{b38} while in the second case the incoming wave interfere destructively with itself between the two interfaces  defining the barrier size ($x=0$ and $x=d$) and transmission resonances do not occur $T(\theta\neq 0)\neq 1$. The resonance condition involves the energy $E / U$ , the width of the barrier d and the angle $\theta$ and reads \cite{b38}: $2D\sqrt{(1-2 E / U+(E / U)^2 cos^2 \theta})=\text{integer}, (D\equiv Ud  / 2\pi)$. In Fig.\ref{fig5} we can observe what happen when the incident electrons energy varies within the range $0<E<U $and takes the values above the barrier height $E>U>0 $ for a given barrier width. Fabry-P\'erot resonances are clearly visible for $E / U < 1 / 2$ (see first graph in the right panel). For $E / U>1 / 2 $  a critical angle defined as $\theta_c=arcsin (|U-E| / E )$ determines the limit below which the transmission can be observed. Some specific regions can be recognized (see right panel in Fig.\ref{fig5}):
\begin{itemize}
\item[(i)] $1/2<E/U<1$ in which the wave is transmitted through the barrier with reduced amplitude except $\theta\approx0$  when T=1;
\item[(ii)]$1<E /U<1.5$  which corresponds to the situation of a $nn'n$ junction and the properties of the transmission coefficient are recovered from the previous case as a mirror reflection;
\item[(iii)] $E / U>1.5$ where the existence of barrier manifests itself in oscillations of transmission coefficient until the high energy limit is reached. 
\end{itemize}

\begin{figure}
\includegraphics[clip, trim=2cm 16cm 2cm 2cm,scale=0.9]{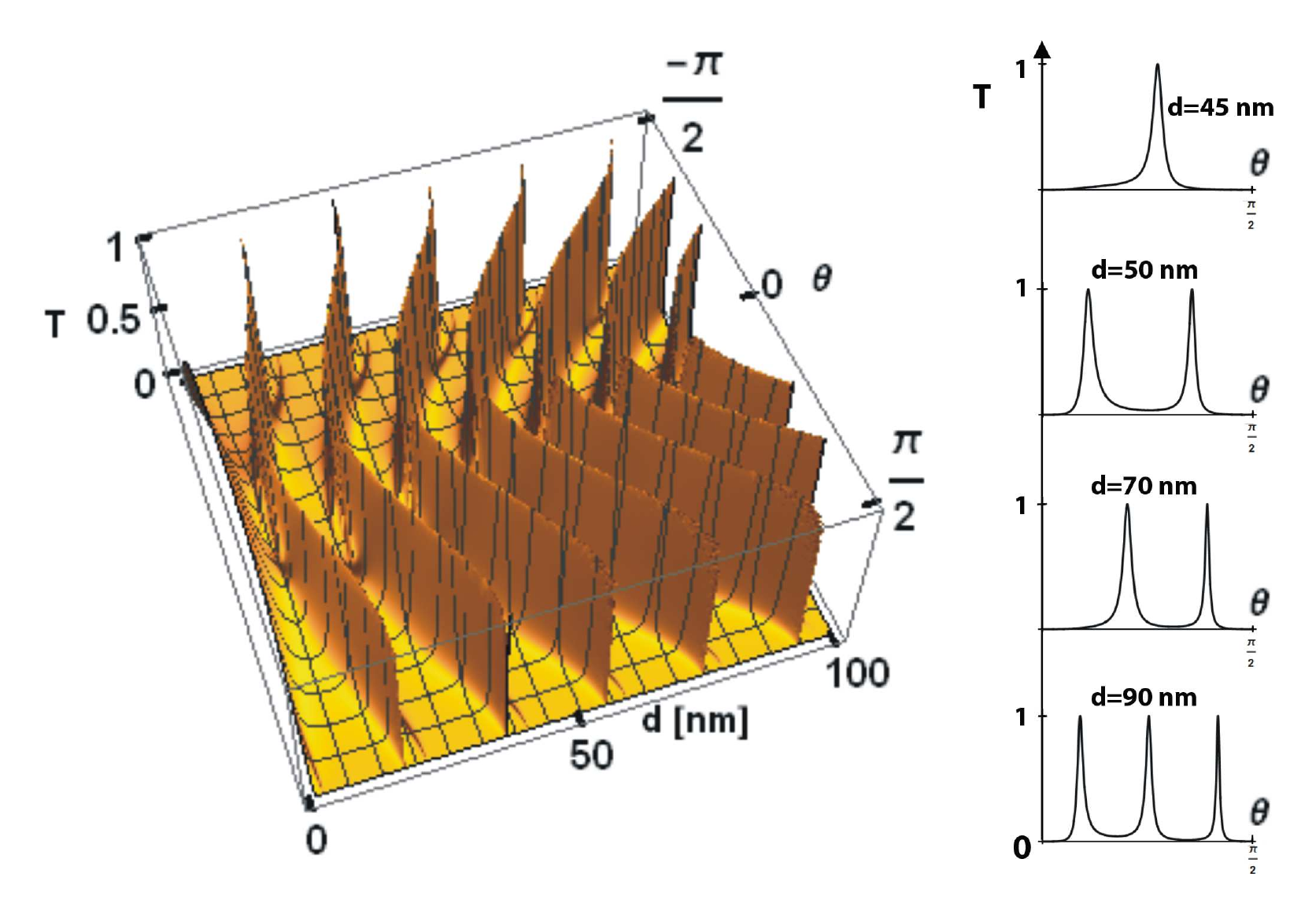}
\caption{\label{fig6}(Color online)  Transmission probability as a function of the incidence angle θ and the barrier width for a fixed energy of incident electrons $(E =$ 17meV) below the barrier height $(U =$ 50meV) calculated for one of the effective BLG existing in 5GL/h-BN/MLG system (black dashed curves in Fig.\ref{fig1}e). Right panel shows the cross-sections for four barrier widths indicating the behavior of "magic angles".  }
\end{figure}

\begin{figure}
\includegraphics[clip, trim=2cm 16.5cm 2cm 2cm,scale=0.9]{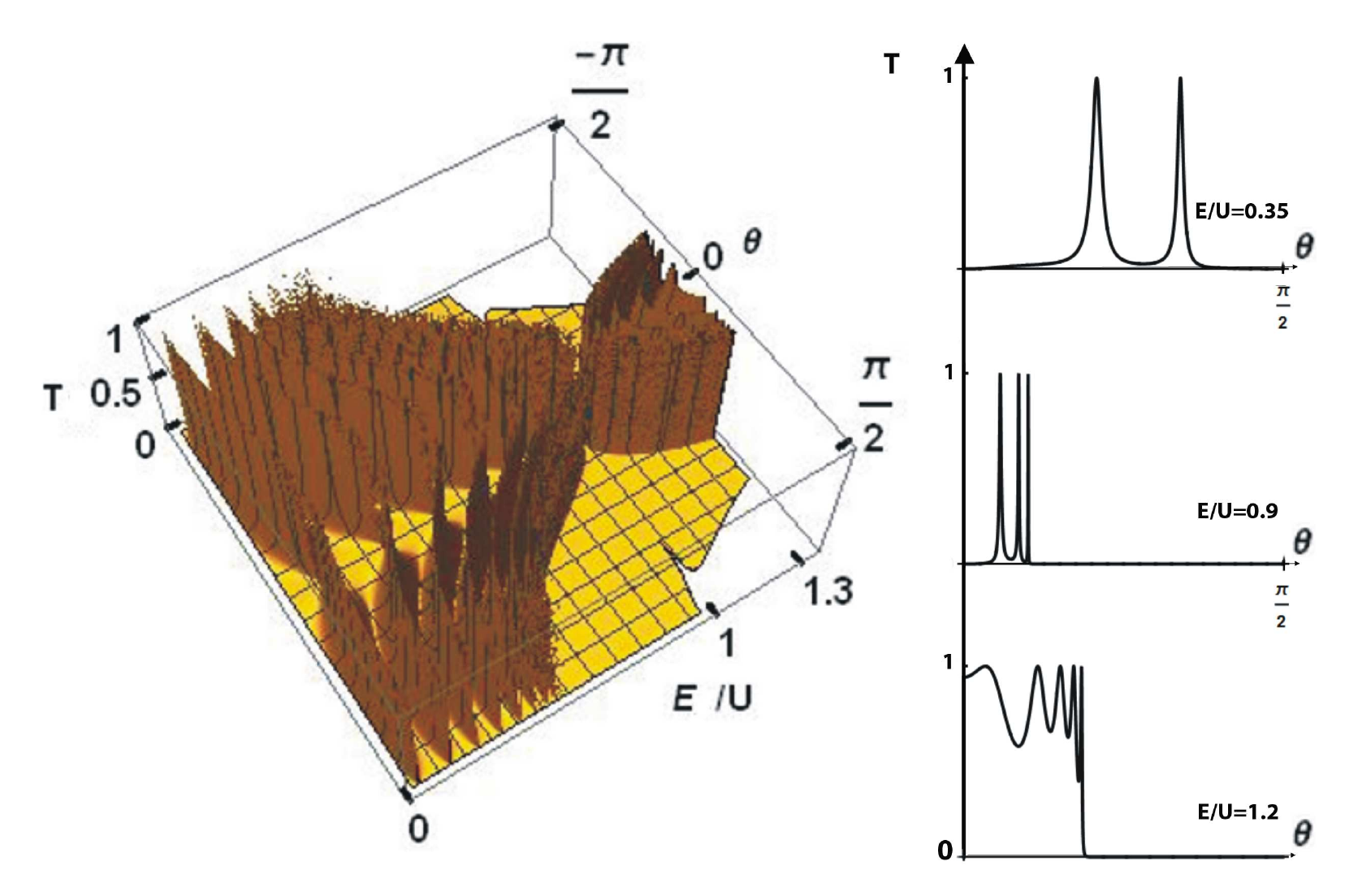}
\caption{\label{fig7}(Color online)  Transmission probability as a function of the incidence angle $\theta$ and the dimensionless energy $E / U$ ($U =$ 50meV) for a fixed width of the barrier ($d =$ 100nm) calculated for one of the effective BLG existing in 5GL/h-BN/MLG system (black dashed curves in Fig.\ref{fig1}e). Right panel shows the cross-sections for three characteristic regions. }
\end{figure}

Electrons in monolayer graphene act like massless spin-1/2 Dirac fermions which results in Klein tunneling with unit transmission for normal incident electrons at a pn junction regardless of barrier height (see Fig.\ref{fig4} and \ref{fig5}). In contrast, bilayer graphene electrons act like parabolic spin-1 systems with perfect reflection for normal incidence and this phenomenon is known as anti-Klein tunneling. It is well visible in Fig.\ref{fig6} were we show the dependence of transmission coefficient on electrons incidence angle and barrier width for the fixed energy of the incident electrons below the barrier height $(0<E<U)$ in the case of one of the effective bilayers found in 5GL/h-BN/MLG system (see black dashed curves in Fig.\ref{fig1}e and Table \ref{tab1}). Other characteristic features known from numerical simulations \cite{b2,b27} are connected with "magic angles" in the spectrum, at which the total transmission is observed (see right panels in Fig.\ref{fig6} and \ref{fig7}). Recently, an explanation of the existence of non-zero "magic angles" with $100 \%$ transmission in the case of symmetric potential barrier, as well as of their almost-survival for slightly asymmetric barrier, has been proposed \cite{b39}. In the present numerical analysis, we can track the behavior of "magic angles" with respect to the barrier width (Fig.\ref{fig6}) and to the energy of the incident electrons (Fig.\ref{fig7}). The number of "magic angles" depends on the energy of the incident electrons while it does not depend on the barrier width. However, their location in the $T(\theta)$ spectrum depends on both parameters. It is worth noting that chiral tunneling presented in Fig.\ref{fig6} and \ref{fig7} can be observed only for the values of parameters $E$ and $U$ from the range $(0,\gamma_{eff} )$. Larger values require higher bands to be considered \cite{b40}.

\subsection{Graphene h-BN hybrid systems supported by h-BN }

Analysis presented in previous subsection reveals that graphene multilayer properties can be tuned by inserting the h-BN layer into the system in the appropriate way. Thus, it seems to be interesting for electronic applications. However, in order to construct nanoelectronic devices one need a support and a natural choice is h-BN.\\
Partition rule indicates unambiguously that introducing the supporting h-BN layer into previously discussed systems eliminates the possibility of preserving graphene monolayer segments. Hybrids of $N_1$/h-BN/$N_2$/h-BN type decouple depending on parity of graphene layers $N_1$ and $N_2$ according to the following scheme:  
\begin{itemize}
\item[(i)] $N_1$ and $N_2$ even - $N/2$  BLG plus $2$ h-BN; 
\item[(ii)] $N_1$ and $N_2$ odd - $(N-2)/2$ BLG plus $2$ MLG/h-BN;
\item[(iii)]$N_1$ even ($N_2$ odd) and  $N_2$ odd ($N_1$ even)- $(N-1)/2$ BLG plus MLG/h-BN plus h-BN.
\end{itemize}

In order to show the volume of quantitative changes induced by h-BN support we present in Table II the effective parameters for 5GL/h-BN/MLG/h-BN system.

\begin{figure}
\includegraphics[clip, trim=2cm 16.5cm 2cm 2cm,scale=0.9]{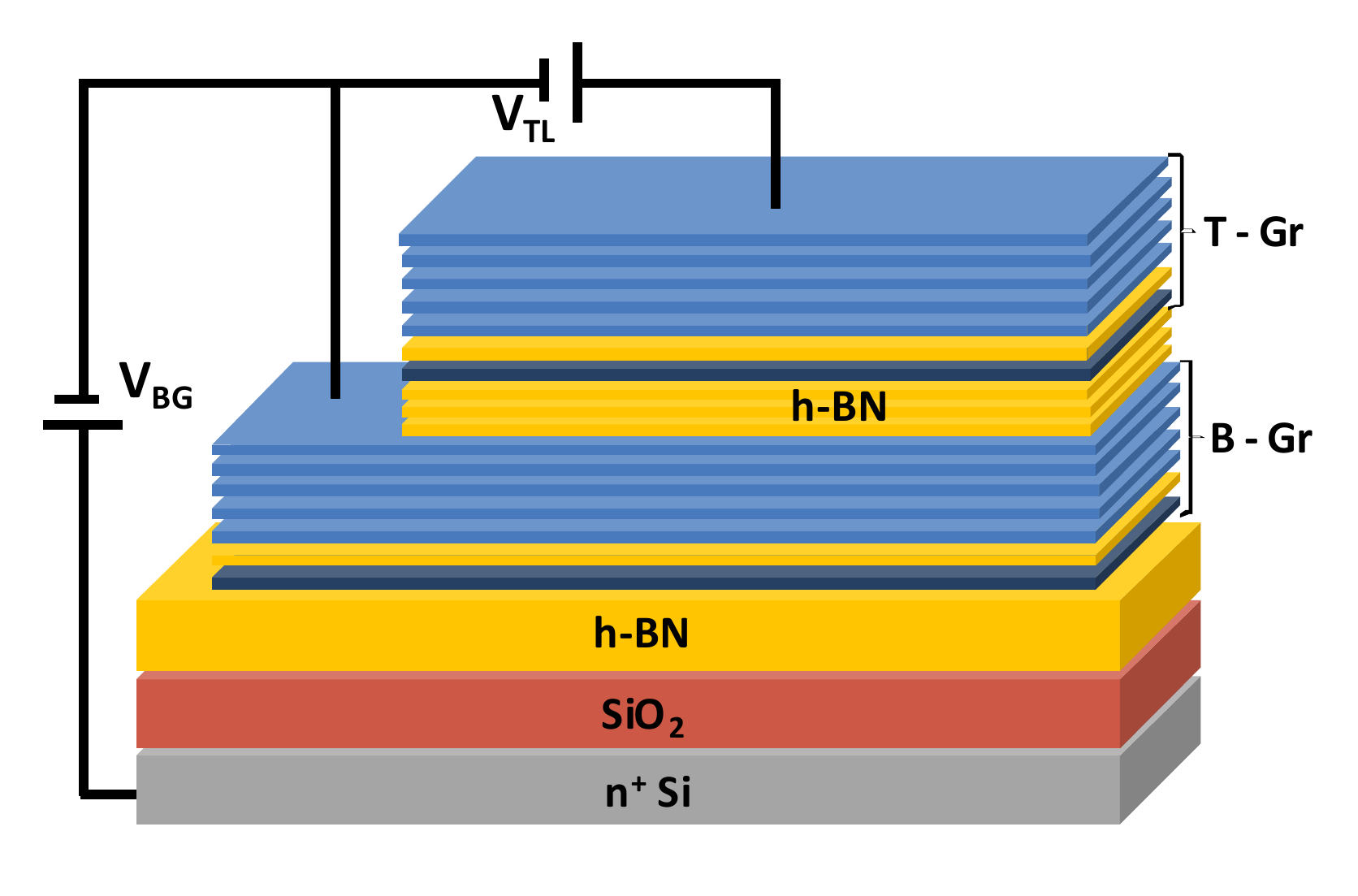}
\caption{\label{fig8}(Color online)  Schematic of the device structure and biasing scheme in analogy to \cite{b51}. The two hybrid electrode layers are separated by a few layer h-BN spacer and supported by thick h-BN. The heavily n-type doped silicon substrate can be used as the back gate.}
\end{figure}

\begin{table}
\caption{\label{tab2}Effective parameters $(\gamma_{\text{\it eff}},\varepsilon_{\text{\it eff}})$ and energy gaps $\Delta_g$ induced by h-BN in 5GL/h-BN/MLG/h-BN system.}
\begin{center}
\begin{tabular}{|c|c|r|r|r|}
\hline
Original multilayer & After decoupling & 
\multicolumn{1}{|c|}{$\gamma_{\text{\it eff}}$ [eV]} &
\multicolumn{1}{|c|}{$\varepsilon_{\text{\it eff}}$ [eV]} &
\multicolumn{1}{|c|}{$\Delta_g$ [eV]} \\
\hline\hline
\multirow{4}{*}{5GL/h-BN/MLG/h-BN} &
BLG&0.675&-0.0016&{}\\
\cline{2-5}	
&BLG	&0.389&-0.0047&\\
\cline{2-5}
&MLG/h-BN&0.098 &3.4033 &0.0028\\
\cline{2-5}
&MLG/h-BN&0.367&3.3250&0.0400\\
\hline
\end{tabular}
\end{center}
\end{table}
Comparison with effective parameters for chiral decomposition of 5GL/h-BN/MLG system (see Table \ref{tab1}) reveals that two bilayer subsystems remain unchanged while "old" MLG/h-BN subsystem is modified. It is interesting that in "new" MLG/h-BN subsystem, the h-BN induced changes in MLG are very small compared to the graphene multilayer system supported by h-BN (see Table I). This rule applies to all hybrid systems of this kind in particular to the one composed of alternating graphene and h-BN monolayers. The band gap of 18 meV opens in MLG/h-BN while the band gap in MLG/h-BN/MLG/h-BN is equal 7 meV (48 meV in second subsystem). Thus, Dirac cone is better protected in this kind of hybrid systems than in graphene supported by h-BN.\\ 
Progress in manufacturing techniques for heterostructures based on graphene and h-BN \cite{b41} $\div$ \cite{b44} has facilitated the study of this hybrid structures for various applications \cite{b45, b46}. One such application is the tunnel field-effect transistor and in particular interlayer tunnel field-effect transistor (ITEFT) \cite{b47}$\div$\cite{b50}.  Quite recently, Sangwoo Kang et al. \cite{b51} explored various combinations of graphene and interlayer hBN thicknesses in order to experimentally study the effects of electrode layer band structure on the characteristics of ITFET's. They studied double bilayer, trilayer, quadlayer, and both Bernal-(ABA)-stacked and rhombohedral-(ABC)-stacked pentalayer graphene as the electrode layer with varying hBN tunnel barrier thicknesses. They found that the differing band structures for the multilayer graphene with varying thicknesses cause significant changes in the interlayer current-voltage characteristics and that certain band structures are more preferable than others in obtaining higher peak-to-valley current ratio. In this context, for electrode layers composed of graphene multilayers with an odd number of layers, constructions with hybrid (see Fig.\ref{fig8}) could probably improve device performance. A separate problem is the ratio of the benefits to the amount of effort involved in preparation of such subtle hybrids. One has to remember, however that heterostructures assembled with atomic layer precision have been already reported \cite{b42} setting the path for experimental work on sophisticated nanoelectronic devices.

\section{Summary and final remarks}

In summary, we studied the electronic properties of multilayer graphene tuned by the appropriate placement of h-BN layers within graphene stacks.  We have turned our attention to the Bernal stacking systems which chiral nature allows decomposition into bilayer systems plus monolayer when the total number of layers is odd. Chiral decomposition procedure \cite{b27} used here for the hybrid multilayer systems reveals a mechanism of compensation against the imposed asymmetry by h-BN layers (see Table \ref{tab1} and \ref{tab2}). Detailed quantitative analysis carried out in this work shows how the h-BN layers can promote specific graphene properties. First of all, a pure graphene monolayer properties are preserved in graphene/h-BN/graphene hybrids of type $N_1$-odd/h-BN/$N_2$-odd. If supported by h-BN only a tiny gap is opened which proves that Dirac cone is better protected in this kind of hybrid systems than in graphene supported by h-BN. This result can be useful in construction of interlayer tunnel-field-effect transistor \cite{b51} where graphene multilayers can be replaced with appropriate hybrids (see Fig.\ref{fig8}). Next, all types of hybrids give the possibility of creating identical bilayers in one system which can be useful when developing bilayer graphene "pseudospintronics" based on the concept of layer pseudospin \cite{b29}.\\
As an example, we provide a pedagogical overview about chiral tunneling in graphene subsystems present in hybrid with one h-BN layer. We discuss then chiral tunneling, Klein tunneling for MLG and anti-Klein tunneling for BLG, analyzing dependence of the transmission coefficients on the angle of incidence, the energy of incident electrons and the width of the potential barrier. We have established the parameter ranges for which the characteristic features in the spectrum are observed, such as Fabry-P\'erot resonances in the case of MLG and "magic angles" in the case of BLG.\\
Based on the results presented in this paper, it turns out that h-BN layers can be treated not only as an insulating undercoat but also as a promoter of graphene multilayer properties. Therefore, taking into account the effects not included in these model considerations would be an interesting future problem. It would be desirable to extend the present analysis to incommensurate hybrid structures not only graphene/h-BN type but also graphene/h-BN mixed with 2D semiconductor.

\appendix
\numberwithin{equation}{section}

\section{Appendix }

The effective Hamiltonians derived for the hybrid systems discussed in subsection 3.1. 

\begin{align}
\label{e21}
&H_{4GL/h-BN}=     \nonumber \\
& \left(  
\begin{smallmatrix}   
\begin{smallmatrix}
\varepsilon_{1,eff} & -\gamma_0pe^{-i\phi} & 0& \gamma_{1,eff} \\    
-\gamma_0pe^{i\phi} & 0 & 0 & 0 \\   
 0 & 0 & 0 & -\gamma_0pe^{-i\phi} \\    
 \gamma_{1,eff} & 0  & -\gamma_0pe^{i\phi} & \varepsilon_{1,eff} \\
 \end{smallmatrix} 
 &   
 \begin{smallmatrix}
 0 \;\;\;\;\;\;\;\;\;\;\;& 0\;\;\;\;\;\;\;\;\;\;\; & 0\;\;\;\;\;\;\;\;\;\;\; & 0\\
  0\;\;\;\;\;\;\;\;\;\;\; & 0\;\;\;\;\;\;\;\;\;\;\; & 0\;\;\;\;\;\;\;\;\;\;\; & 0\\
   0 \;\;\;\;\;\;\;\;\;\;\;& 0\;\;\;\;\;\;\;\;\;\;\; & 0\;\;\;\;\;\;\;\;\;\;\;& 0\\
    0\;\;\;\;\;\;\;\;\;\;\; & 0\;\;\;\;\;\;\;\;\;\;\; & 0\;\;\;\;\;\;\;\;\;\;\;& 0
   \end{smallmatrix}      
   & \begin{smallmatrix}
   0\;\;\;\;\;\;\;\;\;\;\;&0\\
   0\;\;\;\;\;\;\;\;\;\;\;&0\\
   0\;\;\;\;\;\;\;\;\;\;\;&0\\
   0\;\;\;\;\;\;\;\;\;\;\;&0
  \end{smallmatrix}\\
   \begin{smallmatrix}
 0\;\;\;\;\;\;\;\;\;\;\; & 0 \;\;\;\;\;\;\;\;\;\;\;& 0 \;\;\;\;\;\;\;\;\;\;\;& 0\\
  0 \;\;\;\;\;\;\;\;\;\;\;& 0 \;\;\;\;\;\;\;\;\;\;\;& 0\;\;\;\;\;\;\;\;\;\;\;& 0\\
   0 \;\;\;\;\;\;\;\;\;\;\;& 0\;\;\;\;\;\;\;\;\;\;\; & 0\;\;\;\;\;\;\;\;\;\;\; & 0\\
    0\;\;\;\;\;\;\;\;\;\;\; & 0\;\;\;\;\;\;\;\;\;\;\; & 0\;\;\;\;\;\;\;\;\;\;\; & 0
   \end{smallmatrix}   &   
\begin{smallmatrix}
\varepsilon_{2,eff} & -\gamma_0pe^{-i\phi} & 0& \gamma_{2,eff} \\    
-\gamma_0pe^{i\phi} & 0 & 0 & 0 \\   
 0 & 0 & 0 & -\gamma_0pe^{-i\phi} \\    
 \gamma_{2,eff} & 0  & -\gamma_0pe^{i\phi} & \varepsilon_{2,eff} \\
 \end{smallmatrix}       
  & \begin{smallmatrix}
   0\;\;\;\;\;\;\;\;\;\;\;&0\\
   0\;\;\;\;\;\;\;\;\;\;\;&0\\
   0\;\;\;\;\;\;\;\;\;\;\;&0\\
   0\;\;\;\;\;\;\;\;\;\;\;&0
  \end{smallmatrix}\\
  \begin{smallmatrix}
   0\;\;\;\;\;\;\;\;\;\;\;&0\;\;\;\;\;\;\;\;\;\;\;&0\;\;\;\;\;\;\;\;\;\;\;&0\\
   0\;\;\;\;\;\;\;\;\;\;\;&0\;\;\;\;\;\;\;\;\;\;\;&0\;\;\;\;\;\;\;\;\;\;\;&0
  \end{smallmatrix} & \begin{smallmatrix}
   0\;\;\;\;\;\;\;\;\;\;\;&0\;\;\;\;\;\;\;\;\;\;\;&0\;\;\;\;\;\;\;\;\;\;\;&0\\
   0\;\;\;\;\;\;\;\;\;\;\;&0\;\;\;\;\;\;\;\;\;\;\;&0\;\;\;\;\;\;\;\;\;\;\;&0
  \end{smallmatrix} &  \begin{smallmatrix}
   \varepsilon_{\mathrm{N}eff}&-\gamma'_0pe^{-i\phi}\\\\
   -\gamma'_0pe^{-i\phi}&\varepsilon_{\mathrm{B}}
  \end{smallmatrix}
  \end{smallmatrix}
\right)  
\end{align}

\begin{landscape}
\begin{align}
\label{e22}
&H_{4GL/h-BN/MLG}=     \nonumber \\
& \left(  
\begin{smallmatrix}   
\begin{smallmatrix}
\varepsilon_{1,eff} & -\gamma_0pe^{-i\phi} & 0& \gamma_{1,eff} \\    
-\gamma_0pe^{i\phi} & 0 & 0 & 0 \\   
 0 & 0 & 0 & -\gamma_0pe^{-i\phi} \\    
 \gamma_{1,eff} & 0  & -\gamma_0pe^{i\phi} & \varepsilon_{1,eff} \\
 \end{smallmatrix} 
 &   
 \begin{smallmatrix}
 0\;\;\;\;\;\;\;\;\;\;\; & 0\;\;\;\;\;\;\;\;\;\;\;& 0\;\;\;\;\;\;\;\;\;\;\;& 0\\
  0\;\;\;\;\;\;\;\;\;\;\; & 0\;\;\;\;\;\;\;\;\;\;\; & 0\;\;\;\;\;\;\;\;\;\;\; & 0\\
   0\;\;\;\;\;\;\;\;\;\;\; & 0 \;\;\;\;\;\;\;\;\;\;\;&0 \;\;\;\;\;\;\;\;\;\;\; & 0\\
    \gamma_1\;\;\;\;\;\;\;\;\;\;\; & 0 \;\;\;\;\;\;\;\;\;\;\;& 0\;\;\;\;\;\;\;\;\;\;\; & 0
   \end{smallmatrix}      
   & \begin{smallmatrix}
   0\;\;\;\;\;\;\;\;\;\;\; & 0\;\;\;\;\;\;\;\;\;\;\;& 0\;\;\;\;\;\;\;\;\;\;\;& 0\\
  0\;\;\;\;\;\;\;\;\;\;\; & 0\;\;\;\;\;\;\;\;\;\;\; & 0\;\;\;\;\;\;\;\;\;\;\; & 0\\
   0\;\;\;\;\;\;\;\;\;\;\; & 0 \;\;\;\;\;\;\;\;\;\;\;&0 \;\;\;\;\;\;\;\;\;\;\; & 0\\
    0\;\;\;\;\;\;\;\;\;\;\; & 0 \;\;\;\;\;\;\;\;\;\;\;& 0\;\;\;\;\;\;\;\;\;\;\; & 0
  \end{smallmatrix}  \\
   \begin{smallmatrix}
 0\;\;\;\;\;\;\;\;\;\;\; & 0\;\;\;\;\;\;\;\;\;\;\;& 0\;\;\;\;\;\;\;\;\;\;\;& \gamma_1\\
  0\;\;\;\;\;\;\;\;\;\;\; & 0\;\;\;\;\;\;\;\;\;\;\; & 0\;\;\;\;\;\;\;\;\;\;\; & 0\\
   0\;\;\;\;\;\;\;\;\;\;\; & 0 \;\;\;\;\;\;\;\;\;\;\;&0 \;\;\;\;\;\;\;\;\;\;\; & 0\\
   0\;\;\;\;\;\;\;\;\;\;\; & 0 \;\;\;\;\;\;\;\;\;\;\;& 0\;\;\;\;\;\;\;\;\;\;\; & 0
   \end{smallmatrix}   &   
\begin{smallmatrix}
\varepsilon_{2,eff} & -\gamma_0pe^{-i\phi} & 0& \gamma_{2,eff} \\    
-\gamma_0pe^{i\phi} & 0 & 0 & 0 \\   
 0 & 0 & 0 & -\gamma_0pe^{-i\phi} \\    
 \gamma_{2,eff} & 0  & -\gamma_0pe^{i\phi} & \varepsilon_{2,eff} \\
 \end{smallmatrix}       
  & \begin{smallmatrix}
   0\;\;\;\;\;\;\;\;\;\;\; & 0\;\;\;\;\;\;\;\;\;\;\;& 0\;\;\;\;\;\;\;\;\;\;\;& 0\\
  0\;\;\;\;\;\;\;\;\;\;\; & 0\;\;\;\;\;\;\;\;\;\;\; & 0\;\;\;\;\;\;\;\;\;\;\; & 0\\
   0\;\;\;\;\;\;\;\;\;\;\; & 0 \;\;\;\;\;\;\;\;\;\;\;&0 \;\;\;\;\;\;\;\;\;\;\; & 0\\
    0\;\;\;\;\;\;\;\;\;\;\; & 0 \;\;\;\;\;\;\;\;\;\;\;& 0\;\;\;\;\;\;\;\;\;\;\; & 0
  \end{smallmatrix}\\
  \begin{smallmatrix}
   0\;\;\;\;\;\;\;\;\;\;\; & 0\;\;\;\;\;\;\;\;\;\;\;& 0\;\;\;\;\;\;\;\;\;\;\;& 0\\
  0\;\;\;\;\;\;\;\;\;\;\; & 0\;\;\;\;\;\;\;\;\;\;\; & 0\;\;\;\;\;\;\;\;\;\;\; & 0\\
   0\;\;\;\;\;\;\;\;\;\;\; & 0 \;\;\;\;\;\;\;\;\;\;\;&0 \;\;\;\;\;\;\;\;\;\;\; & 0\\
    0\;\;\;\;\;\;\;\;\;\;\; & 0 \;\;\;\;\;\;\;\;\;\;\;& 0\;\;\;\;\;\;\;\;\;\;\; & 0
  \end{smallmatrix} & \begin{smallmatrix}
   0\;\;\;\;\;\;\;\;\;\;\; & 0\;\;\;\;\;\;\;\;\;\;\;& 0\;\;\;\;\;\;\;\;\;\;\;& 0\\
  0\;\;\;\;\;\;\;\;\;\;\; & 0\;\;\;\;\;\;\;\;\;\;\; & 0\;\;\;\;\;\;\;\;\;\;\; & 0\\
   0\;\;\;\;\;\;\;\;\;\;\; & 0 \;\;\;\;\;\;\;\;\;\;\;&0 \;\;\;\;\;\;\;\;\;\;\; & 0\\
    0\;\;\;\;\;\;\;\;\;\;\; & 0 \;\;\;\;\;\;\;\;\;\;\;& 0\;\;\;\;\;\;\;\;\;\;\; & 0
  \end{smallmatrix} &  \begin{smallmatrix}
   \varepsilon_{\mathrm{N}eff}&-\gamma'_0pe^{-i\phi}&0& \varepsilon_{\mathrm{C-N}eff}\\
   -\gamma'_0pe^{-i\phi}&\varepsilon_{\mathrm{B}}&0&0\\
   0&0&0&-\gamma_0pe^{-i\phi}\\
   \varepsilon_{\mathrm{C-N}eff}&0&-\gamma_0pe^{i\phi}&0
  \end{smallmatrix}
  \end{smallmatrix}
\right)  
\end{align}

\begin{align}
\label{e23}
&H_{TLG/h-BN/BLG}=     \nonumber \\
& \left(  
\begin{smallmatrix}   
\begin{smallmatrix}
\varepsilon_{1,eff} & -\gamma_0pe^{-i\phi} & 0& \gamma_{1,eff} \\    
-\gamma_0pe^{i\phi} & 0 & 0 & 0 \\   
 0 & 0 & 0 & -\gamma_0pe^{-i\phi} \\    
 \gamma_{1,eff} & 0  & -\gamma_0pe^{i\phi} & \varepsilon_{1,eff} \\
 \end{smallmatrix} 
 &   
 \begin{smallmatrix}
  0\;\;\;\;\;\;\;\;\;\;\; & 0\;\;\;\;\;\;\;\;\;\;\;& 0\;\;\;\;\;\;\;\;\;\;\;& 0\\
  0\;\;\;\;\;\;\;\;\;\;\; & 0\;\;\;\;\;\;\;\;\;\;\; & 0\;\;\;\;\;\;\;\;\;\;\; & 0\\
   0\;\;\;\;\;\;\;\;\;\;\; & 0 \;\;\;\;\;\;\;\;\;\;\;&0 \;\;\;\;\;\;\;\;\;\;\; & 0\\
    0\;\;\;\;\;\;\;\;\;\;\; & 0 \;\;\;\;\;\;\;\;\;\;\;& 0\;\;\;\;\;\;\;\;\;\;\; & 0
   \end{smallmatrix}      
   & \begin{smallmatrix}
    0\;\;\;\;\;\;\;\;\;\;\; & 0\;\;\;\;\;\;\;\;\;\;\;& 0\;\;\;\;\;\;\;\;\;\;\;& 0\\
  0\;\;\;\;\;\;\;\;\;\;\; & 0\;\;\;\;\;\;\;\;\;\;\; & 0\;\;\;\;\;\;\;\;\;\;\; & 0\\
   0\;\;\;\;\;\;\;\;\;\;\; & 0 \;\;\;\;\;\;\;\;\;\;\;&0 \;\;\;\;\;\;\;\;\;\;\; & 0\\
    0\;\;\;\;\;\;\;\;\;\;\; & 0 \;\;\;\;\;\;\;\;\;\;\;& 0\;\;\;\;\;\;\;\;\;\;\; & 0
  \end{smallmatrix}  \\
   \begin{smallmatrix}
  0\;\;\;\;\;\;\;\;\;\;\; & 0\;\;\;\;\;\;\;\;\;\;\;& 0\;\;\;\;\;\;\;\;\;\;\;& 0\\
  0\;\;\;\;\;\;\;\;\;\;\; & 0\;\;\;\;\;\;\;\;\;\;\; & 0\;\;\;\;\;\;\;\;\;\;\; & 0\\
   0\;\;\;\;\;\;\;\;\;\;\; & 0 \;\;\;\;\;\;\;\;\;\;\;&0 \;\;\;\;\;\;\;\;\;\;\; & 0\\
    0\;\;\;\;\;\;\;\;\;\;\; & 0 \;\;\;\;\;\;\;\;\;\;\;& 0\;\;\;\;\;\;\;\;\;\;\; & 0
   \end{smallmatrix}   &   
\begin{smallmatrix}
0 & -\gamma_0pe^{-i\phi} & 0& \gamma_{\mathrm{C-N},eff} \\    
-\gamma_0pe^{i\phi} & 0 & 0 & 0 \\   
 0 & 0 & \varepsilon_{\mathrm{B}} & -\gamma'_0pe^{-i\phi} \\    
 \gamma_{\mathrm{C-N},eff} & 0  & -\gamma'_0pe^{i\phi} & \varepsilon_{\mathrm{N},eff} \\
 \end{smallmatrix}       
  & \begin{smallmatrix}
    0\;\;\;\;\;\;\;\;\;\;\; & 0\;\;\;\;\;\;\;\;\;\;\;& 0\;\;\;\;\;\;\;\;\;\;\;& 0\\
  0\;\;\;\;\;\;\;\;\;\;\; & 0\;\;\;\;\;\;\;\;\;\;\; & 0\;\;\;\;\;\;\;\;\;\;\; & 0\\
   0\;\;\;\;\;\;\;\;\;\;\; & 0 \;\;\;\;\;\;\;\;\;\;\;&0 \;\;\;\;\;\;\;\;\;\;\; & 0\\
    0\;\;\;\;\;\;\;\;\;\;\; & 0 \;\;\;\;\;\;\;\;\;\;\;& 0\;\;\;\;\;\;\;\;\;\;\; & 0
  \end{smallmatrix}\\
  \begin{smallmatrix}
    0\;\;\;\;\;\;\;\;\;\;\; & 0\;\;\;\;\;\;\;\;\;\;\;& 0\;\;\;\;\;\;\;\;\;\;\;& 0\\
  0\;\;\;\;\;\;\;\;\;\;\; & 0\;\;\;\;\;\;\;\;\;\;\; & 0\;\;\;\;\;\;\;\;\;\;\; & 0\\
   0\;\;\;\;\;\;\;\;\;\;\; & 0 \;\;\;\;\;\;\;\;\;\;\;&0 \;\;\;\;\;\;\;\;\;\;\; & 0\\
    0\;\;\;\;\;\;\;\;\;\;\; & 0 \;\;\;\;\;\;\;\;\;\;\;& 0\;\;\;\;\;\;\;\;\;\;\; & 0
  \end{smallmatrix} & \begin{smallmatrix}
    0\;\;\;\;\;\;\;\;\;\;\; & 0\;\;\;\;\;\;\;\;\;\;\;& 0\;\;\;\;\;\;\;\;\;\;\;& 0\\
  0\;\;\;\;\;\;\;\;\;\;\; & 0\;\;\;\;\;\;\;\;\;\;\; & 0\;\;\;\;\;\;\;\;\;\;\; & 0\\
   0\;\;\;\;\;\;\;\;\;\;\; & 0 \;\;\;\;\;\;\;\;\;\;\;&0 \;\;\;\;\;\;\;\;\;\;\; & 0\\
    0\;\;\;\;\;\;\;\;\;\;\; & 0 \;\;\;\;\;\;\;\;\;\;\;& 0\;\;\;\;\;\;\;\;\;\;\; & 0
  \end{smallmatrix} &  \begin{smallmatrix}
   \varepsilon_{2,eff}&-\gamma_0pe^{-i\phi}&0& \gamma_{2,eff}\\
   -\gamma_0pe^{i\phi}&0&0&0\\
   0&0&0&-\gamma_0pe^{-i\phi}\\
  \gamma_{2,eff}&0&-\gamma_0pe^{i\phi}&\varepsilon_{2,eff}
  \end{smallmatrix}
  \end{smallmatrix}
\right)  
\end{align}

\begin{align}
\label{e24}
&H_{4GL/h-BN/BLG}=     \nonumber \\
& \left(  
\begin{smallmatrix}   
\begin{smallmatrix}
\varepsilon_{1,eff} & -\gamma_0pe^{-i\phi} & 0& \gamma_{1,eff} \\    
-\gamma_0pe^{i\phi} & 0 & 0 & 0 \\   
 0 & 0 & 0 & -\gamma_0pe^{-i\phi} \\    
 \gamma_{1,eff} & 0  & -\gamma_0pe^{i\phi} & \varepsilon_{1,eff} \\
 \end{smallmatrix} 
 &   
 \begin{smallmatrix}
  0\;\;\;\;\;\;\;\;\;\;\; & 0\;\;\;\;\;\;\;\;\;\;\;& 0\;\;\;\;\;\;\;\;\;\;\;& 0\\
  0\;\;\;\;\;\;\;\;\;\;\; & 0\;\;\;\;\;\;\;\;\;\;\; & 0\;\;\;\;\;\;\;\;\;\;\; & 0\\
   0\;\;\;\;\;\;\;\;\;\;\; & 0 \;\;\;\;\;\;\;\;\;\;\;&0 \;\;\;\;\;\;\;\;\;\;\; & 0\\
    0\;\;\;\;\;\;\;\;\;\;\; & 0 \;\;\;\;\;\;\;\;\;\;\;& 0\;\;\;\;\;\;\;\;\;\;\; & 0
   \end{smallmatrix}
   &  \begin{smallmatrix}
  0\;\;\;\;\;\;\;\;\;\;\; &  0\\
  0\;\;\;\;\;\;\;\;\;\;\; &  0\\
   0\;\;\;\;\;\;\;\;\;\;\; &  0\\
    0\;\;\;\;\;\;\;\;\;\;\; &   0
   \end{smallmatrix}    
   & \begin{smallmatrix}
    0\;\;\;\;\;\;\;\;\;\;\; & 0\;\;\;\;\;\;\;\;\;\;\;& 0\;\;\;\;\;\;\;\;\;\;\;& 0\\
  0\;\;\;\;\;\;\;\;\;\;\; & 0\;\;\;\;\;\;\;\;\;\;\; & 0\;\;\;\;\;\;\;\;\;\;\; & 0\\
   0\;\;\;\;\;\;\;\;\;\;\; & 0 \;\;\;\;\;\;\;\;\;\;\;&0 \;\;\;\;\;\;\;\;\;\;\; & 0\\
    0\;\;\;\;\;\;\;\;\;\;\; & 0 \;\;\;\;\;\;\;\;\;\;\;& 0\;\;\;\;\;\;\;\;\;\;\; & 0
  \end{smallmatrix}  \\
   \begin{smallmatrix}
  0\;\;\;\;\;\;\;\;\;\;\; & 0\;\;\;\;\;\;\;\;\;\;\;& 0\;\;\;\;\;\;\;\;\;\;\;& 0\\
  0\;\;\;\;\;\;\;\;\;\;\; & 0\;\;\;\;\;\;\;\;\;\;\; & 0\;\;\;\;\;\;\;\;\;\;\; & 0\\
   0\;\;\;\;\;\;\;\;\;\;\; & 0 \;\;\;\;\;\;\;\;\;\;\;&0 \;\;\;\;\;\;\;\;\;\;\; & 0\\
    0\;\;\;\;\;\;\;\;\;\;\; & 0 \;\;\;\;\;\;\;\;\;\;\;& 0\;\;\;\;\;\;\;\;\;\;\; & 0
   \end{smallmatrix}   &   
\begin{smallmatrix}
\varepsilon_{2,eff} & -\gamma_0pe^{-i\phi} & 0& \gamma_{2,eff} \\    
-\gamma_0pe^{i\phi} & 0 & 0 & 0 \\   
 0 & 0 & 0 & -\gamma_0pe^{-i\phi} \\    
 \gamma_{2,eff} & 0  & -\gamma_0pe^{i\phi} & \varepsilon_{2,eff} \\
 \end{smallmatrix}  
 &  \begin{smallmatrix}
  0\;\;\;\;\;\;\;\;\;\;\; &  0\\
  0\;\;\;\;\;\;\;\;\;\;\; &  0\\
   0\;\;\;\;\;\;\;\;\;\;\; &  0\\
    0\;\;\;\;\;\;\;\;\;\;\; &   0
   \end{smallmatrix}         
  & \begin{smallmatrix}
    0\;\;\;\;\;\;\;\;\;\;\; & 0\;\;\;\;\;\;\;\;\;\;\;& 0\;\;\;\;\;\;\;\;\;\;\;& 0\\
  0\;\;\;\;\;\;\;\;\;\;\; & 0\;\;\;\;\;\;\;\;\;\;\; & 0\;\;\;\;\;\;\;\;\;\;\; & 0\\
   0\;\;\;\;\;\;\;\;\;\;\; & 0 \;\;\;\;\;\;\;\;\;\;\;&0 \;\;\;\;\;\;\;\;\;\;\; & 0\\
    0\;\;\;\;\;\;\;\;\;\;\; & 0 \;\;\;\;\;\;\;\;\;\;\;& 0\;\;\;\;\;\;\;\;\;\;\; & 0
  \end{smallmatrix}\\
  \begin{smallmatrix}
   0\;\;\;\;\;\;\;\;\;\;\; & 0 \;\;\;\;\;\;\;\;\;\;\;&0 \;\;\;\;\;\;\;\;\;\;\; & 0\\
    0\;\;\;\;\;\;\;\;\;\;\; & 0 \;\;\;\;\;\;\;\;\;\;\;& 0\;\;\;\;\;\;\;\;\;\;\; & 0
  \end{smallmatrix}&
  \begin{smallmatrix}
   0\;\;\;\;\;\;\;\;\;\;\; & 0 \;\;\;\;\;\;\;\;\;\;\;&0 \;\;\;\;\;\;\;\;\;\;\; & 0\\
    0\;\;\;\;\;\;\;\;\;\;\; & 0 \;\;\;\;\;\;\;\;\;\;\;& 0\;\;\;\;\;\;\;\;\;\;\; & 0
  \end{smallmatrix}&
  \begin{smallmatrix} 
   \varepsilon_{\mathrm{N},eff} & -\gamma'_0pe^{-i\phi} \\
    -\gamma'_0pe^{i\phi} & \varepsilon_{\mathrm{B}} 
  \end{smallmatrix}
  &\begin{smallmatrix}
   0\;\;\;\;\;\;\;\;\;\;\; & 0 \;\;\;\;\;\;\;\;\;\;\;&0 \;\;\;\;\;\;\;\;\;\;\; & 0\\
    0\;\;\;\;\;\;\;\;\;\;\; & 0 \;\;\;\;\;\;\;\;\;\;\;& 0\;\;\;\;\;\;\;\;\;\;\; & 0
  \end{smallmatrix}\\ 
  \begin{smallmatrix}
    0\;\;\;\;\;\;\;\;\;\;\; & 0\;\;\;\;\;\;\;\;\;\;\;& 0\;\;\;\;\;\;\;\;\;\;\;& 0\\
  0\;\;\;\;\;\;\;\;\;\;\; & 0\;\;\;\;\;\;\;\;\;\;\; & 0\;\;\;\;\;\;\;\;\;\;\; & 0\\
   0\;\;\;\;\;\;\;\;\;\;\; & 0 \;\;\;\;\;\;\;\;\;\;\;&0 \;\;\;\;\;\;\;\;\;\;\; & 0\\
    0\;\;\;\;\;\;\;\;\;\;\; & 0 \;\;\;\;\;\;\;\;\;\;\;& 0\;\;\;\;\;\;\;\;\;\;\; & 0
  \end{smallmatrix} & \begin{smallmatrix}
    0\;\;\;\;\;\;\;\;\;\;\; & 0\;\;\;\;\;\;\;\;\;\;\;& 0\;\;\;\;\;\;\;\;\;\;\;& 0\\
  0\;\;\;\;\;\;\;\;\;\;\; & 0\;\;\;\;\;\;\;\;\;\;\; & 0\;\;\;\;\;\;\;\;\;\;\; & 0\\
   0\;\;\;\;\;\;\;\;\;\;\; & 0 \;\;\;\;\;\;\;\;\;\;\;&0 \;\;\;\;\;\;\;\;\;\;\; & 0\\
    0\;\;\;\;\;\;\;\;\;\;\; & 0 \;\;\;\;\;\;\;\;\;\;\;& 0\;\;\;\;\;\;\;\;\;\;\; & 0
  \end{smallmatrix} & \begin{smallmatrix}
  0\;\;\;\;\;\;\;\;\;\;\; &  0\\
  0\;\;\;\;\;\;\;\;\;\;\; &  0\\
   0\;\;\;\;\;\;\;\;\;\;\; &  0\\
    0\;\;\;\;\;\;\;\;\;\;\; &   0
   \end{smallmatrix}  & 
  \begin{smallmatrix}
   0&-\gamma_0pe^{-i\phi}&0& 0\\
   -\gamma_0pe^{i\phi}&\varepsilon_{3,eff}&\gamma_{3,eff}&0\\
   0&\gamma_{3,eff}&\varepsilon_{3,eff}&-\gamma_0pe^{-i\phi}\\
  0&0&-\gamma_0pe^{i\phi}&0
  \end{smallmatrix}
  \end{smallmatrix}
\right)  
\end{align}

\begin{align}
\label{e25}
&H_{5GL/h-BN/MLG}=     \nonumber \\
& \left(  
\begin{smallmatrix}   
\begin{smallmatrix}
\varepsilon_{1,eff} & -\gamma_0pe^{-i\phi} & 0& \gamma_{1,eff} \\    
-\gamma_0pe^{i\phi} & 0 & 0 & 0 \\   
 0 & 0 & 0 & -\gamma_0pe^{-i\phi} \\    
 \gamma_{1,eff} & 0  & -\gamma_0pe^{i\phi} & \varepsilon_{1,eff} \\
 \end{smallmatrix} 
 &   
 \begin{smallmatrix}
  0\;\;\;\;\;\;\;\;\;\;\; & 0\;\;\;\;\;\;\;\;\;\;\;& 0\;\;\;\;\;\;\;\;\;\;\;& 0\\
  0\;\;\;\;\;\;\;\;\;\;\; & 0\;\;\;\;\;\;\;\;\;\;\; & 0\;\;\;\;\;\;\;\;\;\;\; & 0\\
   0\;\;\;\;\;\;\;\;\;\;\; & 0 \;\;\;\;\;\;\;\;\;\;\;&0 \;\;\;\;\;\;\;\;\;\;\; & 0\\
    0\;\;\;\;\;\;\;\;\;\;\; & 0 \;\;\;\;\;\;\;\;\;\;\;& 0\;\;\;\;\;\;\;\;\;\;\; & 0
   \end{smallmatrix}    
   & \begin{smallmatrix}
    0\;\;\;\;\;\;\;\;\;\;\; & 0\;\;\;\;\;\;\;\;\;\;\;& 0\;\;\;\;\;\;\;\;\;\;\;& 0\\
  0\;\;\;\;\;\;\;\;\;\;\; & 0\;\;\;\;\;\;\;\;\;\;\; & 0\;\;\;\;\;\;\;\;\;\;\; & 0\\
   0\;\;\;\;\;\;\;\;\;\;\; & 0 \;\;\;\;\;\;\;\;\;\;\;&0 \;\;\;\;\;\;\;\;\;\;\; & 0\\
    0\;\;\;\;\;\;\;\;\;\;\; & 0 \;\;\;\;\;\;\;\;\;\;\;& 0\;\;\;\;\;\;\;\;\;\;\; & 0
  \end{smallmatrix}  &  \begin{smallmatrix}
  0\;\;\;\;\;\;\;\;\;\;\; &  0\\
  0\;\;\;\;\;\;\;\;\;\;\; &  0\\
   0\;\;\;\;\;\;\;\;\;\;\; &  0\\
    0\;\;\;\;\;\;\;\;\;\;\; &   0
   \end{smallmatrix}  \\
   \begin{smallmatrix}
  0\;\;\;\;\;\;\;\;\;\;\; & 0\;\;\;\;\;\;\;\;\;\;\;& 0\;\;\;\;\;\;\;\;\;\;\;& 0\\
  0\;\;\;\;\;\;\;\;\;\;\; & 0\;\;\;\;\;\;\;\;\;\;\; & 0\;\;\;\;\;\;\;\;\;\;\; & 0\\
   0\;\;\;\;\;\;\;\;\;\;\; & 0 \;\;\;\;\;\;\;\;\;\;\;&0 \;\;\;\;\;\;\;\;\;\;\; & 0\\
    0\;\;\;\;\;\;\;\;\;\;\; & 0 \;\;\;\;\;\;\;\;\;\;\;& 0\;\;\;\;\;\;\;\;\;\;\; & 0
   \end{smallmatrix}   &   
\begin{smallmatrix}
\varepsilon_{2,eff} & -\gamma_0pe^{-i\phi} & 0& \gamma_{2,eff} \\    
-\gamma_0pe^{i\phi} & 0 & 0 & 0 \\   
 0 & 0 & 0 & -\gamma_0pe^{-i\phi} \\    
 \gamma_{2,eff} & 0  & -\gamma_0pe^{i\phi} & \varepsilon_{2,eff} \\
 \end{smallmatrix}          
  & \begin{smallmatrix}
    0\;\;\;\;\;\;\;\;\;\;\; & 0\;\;\;\;\;\;\;\;\;\;\;& 0\;\;\;\;\;\;\;\;\;\;\;& 0\\
  0\;\;\;\;\;\;\;\;\;\;\; & 0\;\;\;\;\;\;\;\;\;\;\; & 0\;\;\;\;\;\;\;\;\;\;\; & 0\\
   0\;\;\;\;\;\;\;\;\;\;\; & 0 \;\;\;\;\;\;\;\;\;\;\;&0 \;\;\;\;\;\;\;\;\;\;\; & 0\\
    0\;\;\;\;\;\;\;\;\;\;\; & 0 \;\;\;\;\;\;\;\;\;\;\;& 0\;\;\;\;\;\;\;\;\;\;\; & 0
  \end{smallmatrix}  &  \begin{smallmatrix}
  0\;\;\;\;\;\;\;\;\;\;\; &  0\\
  0\;\;\;\;\;\;\;\;\;\;\; &  0\\
   0\;\;\;\;\;\;\;\;\;\;\; &  0\\
    0\;\;\;\;\;\;\;\;\;\;\; &   0
   \end{smallmatrix} \\ 
  \begin{smallmatrix}
    0\;\;\;\;\;\;\;\;\;\;\; & 0\;\;\;\;\;\;\;\;\;\;\;& 0\;\;\;\;\;\;\;\;\;\;\;& 0\\
  0\;\;\;\;\;\;\;\;\;\;\; & 0\;\;\;\;\;\;\;\;\;\;\; & 0\;\;\;\;\;\;\;\;\;\;\; & 0\\
   0\;\;\;\;\;\;\;\;\;\;\; & 0 \;\;\;\;\;\;\;\;\;\;\;&0 \;\;\;\;\;\;\;\;\;\;\; & 0\\
    0\;\;\;\;\;\;\;\;\;\;\; & 0 \;\;\;\;\;\;\;\;\;\;\;& 0\;\;\;\;\;\;\;\;\;\;\; & 0
  \end{smallmatrix} & \begin{smallmatrix}
    0\;\;\;\;\;\;\;\;\;\;\; & 0\;\;\;\;\;\;\;\;\;\;\;& 0\;\;\;\;\;\;\;\;\;\;\;& 0\\
  0\;\;\;\;\;\;\;\;\;\;\; & 0\;\;\;\;\;\;\;\;\;\;\; & 0\;\;\;\;\;\;\;\;\;\;\; & 0\\
   0\;\;\;\;\;\;\;\;\;\;\; & 0 \;\;\;\;\;\;\;\;\;\;\;&0 \;\;\;\;\;\;\;\;\;\;\; & 0\\
    0\;\;\;\;\;\;\;\;\;\;\; & 0 \;\;\;\;\;\;\;\;\;\;\;& 0\;\;\;\;\;\;\;\;\;\;\; & 0
  \end{smallmatrix}  & 
  \begin{smallmatrix}
   0&-\gamma_0pe^{-i\phi}&0& \gamma_{\mathrm{C-N},eff}\\
   -\gamma_0pe^{-i\phi}&0&0&0\\
   0&0&\varepsilon_{\mathrm{B}}&-\gamma'_0pe^{-i\phi}\\
  \gamma_{\mathrm{C-N},eff}&0&-\gamma'_0pe^{i\phi}&\varepsilon_{\mathrm{N},eff}
  \end{smallmatrix}& \begin{smallmatrix}
  0\;\;\;\;\;\;\;\;\;\;\; &  0\\
  0\;\;\;\;\;\;\;\;\;\;\; &  0\\
   0\;\;\;\;\;\;\;\;\;\;\; &  0\\
    0\;\;\;\;\;\;\;\;\;\;\; &   0
   \end{smallmatrix} \\
   \begin{smallmatrix}
   0\;\;\;\;\;\;\;\;\;\;\; & 0 \;\;\;\;\;\;\;\;\;\;\;&0 \;\;\;\;\;\;\;\;\;\;\; & 0\\
    0\;\;\;\;\;\;\;\;\;\;\; & 0 \;\;\;\;\;\;\;\;\;\;\;& 0\;\;\;\;\;\;\;\;\;\;\; & 0
  \end{smallmatrix}&\begin{smallmatrix}
   0\;\;\;\;\;\;\;\;\;\;\; & 0 \;\;\;\;\;\;\;\;\;\;\;&0 \;\;\;\;\;\;\;\;\;\;\; & 0\\
    0\;\;\;\;\;\;\;\;\;\;\; & 0 \;\;\;\;\;\;\;\;\;\;\;& 0\;\;\;\;\;\;\;\;\;\;\; & 0
  \end{smallmatrix}&\begin{smallmatrix}
   0\;\;\;\;\;\;\;\;\;\;\; & 0 \;\;\;\;\;\;\;\;\;\;\;&0 \;\;\;\;\;\;\;\;\;\;\; & 0\\
    0\;\;\;\;\;\;\;\;\;\;\; & 0 \;\;\;\;\;\;\;\;\;\;\;& 0\;\;\;\;\;\;\;\;\;\;\; & 0
  \end{smallmatrix}&\begin{smallmatrix}
   0 & -\gamma_0pe^{-i\phi}\\
    -\gamma_0pe^{i\phi}  & 0
  \end{smallmatrix}
  \end{smallmatrix}
\right)  
\end{align}

\begin{align}
\label{e26}
&H_{TLG/h-BN/TLG}=     \nonumber \\
& \left(  
\begin{smallmatrix}   
\begin{smallmatrix}
\varepsilon_{1,eff} & -\gamma_0pe^{-i\phi} & 0& \gamma_{1,eff} \\    
-\gamma_0pe^{i\phi} & 0 & 0 & 0 \\   
 0 & 0 & 0 & -\gamma_0pe^{-i\phi} \\    
 \gamma_{1,eff} & 0  & -\gamma_0pe^{i\phi} & \varepsilon_{1,eff} \\
 \end{smallmatrix} 
 &   
 \begin{smallmatrix}
  0\;\;\;\;\;\;\;\;\;\;\; & 0\;\;\;\;\;\;\;\;\;\;\;& 0\;\;\;\;\;\;\;\;\;\;\;& 0\\
  0\;\;\;\;\;\;\;\;\;\;\; & 0\;\;\;\;\;\;\;\;\;\;\; & 0\;\;\;\;\;\;\;\;\;\;\; & 0\\
   0\;\;\;\;\;\;\;\;\;\;\; & 0 \;\;\;\;\;\;\;\;\;\;\;&0 \;\;\;\;\;\;\;\;\;\;\; & 0\\
    0\;\;\;\;\;\;\;\;\;\;\; & 0 \;\;\;\;\;\;\;\;\;\;\;& 0\;\;\;\;\;\;\;\;\;\;\; & 0
   \end{smallmatrix}
   &  \begin{smallmatrix}
  0\;\;\;\;\;\;\;\;\;\;\; &  0\\
  0\;\;\;\;\;\;\;\;\;\;\; &  0\\
   0\;\;\;\;\;\;\;\;\;\;\; &  0\\
    0\;\;\;\;\;\;\;\;\;\;\; &   0
   \end{smallmatrix}    
   & \begin{smallmatrix}
    0\;\;\;\;\;\;\;\;\;\;\; & 0\;\;\;\;\;\;\;\;\;\;\;& 0\;\;\;\;\;\;\;\;\;\;\;& 0\\
  0\;\;\;\;\;\;\;\;\;\;\; & 0\;\;\;\;\;\;\;\;\;\;\; & 0\;\;\;\;\;\;\;\;\;\;\; & 0\\
   0\;\;\;\;\;\;\;\;\;\;\; & 0 \;\;\;\;\;\;\;\;\;\;\;&0 \;\;\;\;\;\;\;\;\;\;\; & 0\\
    0\;\;\;\;\;\;\;\;\;\;\; & 0 \;\;\;\;\;\;\;\;\;\;\;& 0\;\;\;\;\;\;\;\;\;\;\; & 0
  \end{smallmatrix}  \\
   \begin{smallmatrix}
  0\;\;\;\;\;\;\;\;\;\;\; & 0\;\;\;\;\;\;\;\;\;\;\;& 0\;\;\;\;\;\;\;\;\;\;\;& 0\\
  0\;\;\;\;\;\;\;\;\;\;\; & 0\;\;\;\;\;\;\;\;\;\;\; & 0\;\;\;\;\;\;\;\;\;\;\; & 0\\
   0\;\;\;\;\;\;\;\;\;\;\; & 0 \;\;\;\;\;\;\;\;\;\;\;&0 \;\;\;\;\;\;\;\;\;\;\; & 0\\
    0\;\;\;\;\;\;\;\;\;\;\; & 0 \;\;\;\;\;\;\;\;\;\;\;& 0\;\;\;\;\;\;\;\;\;\;\; & 0
   \end{smallmatrix}   &   
\begin{smallmatrix}
0 & -\gamma_0pe^{-i\phi} & 0& \gamma_{\mathrm{C-N},eff} \\    
-\gamma_0pe^{i\phi} & 0 & 0 & 0 \\   
 0 & 0 & \varepsilon_{\mathrm{B}} & -\gamma'_0pe^{-i\phi} \\    
 \gamma_{\mathrm{C-N},eff} & 0  & -\gamma'_0pe^{i\phi} & \varepsilon_{\mathrm{N},eff} \\
 \end{smallmatrix}  
 &  \begin{smallmatrix}
  0\;\;\;\;\;\;\;\;\;\;\; &  0\\
  0\;\;\;\;\;\;\;\;\;\;\; &  0\\
   0\;\;\;\;\;\;\;\;\;\;\; &  0\\
    0\;\;\;\;\;\;\;\;\;\;\; &   0
   \end{smallmatrix}         
  & \begin{smallmatrix}
    0\;\;\;\;\;\;\;\;\;\;\; & 0\;\;\;\;\;\;\;\;\;\;\;& 0\;\;\;\;\;\;\;\;\;\;\;& 0\\
  0\;\;\;\;\;\;\;\;\;\;\; & 0\;\;\;\;\;\;\;\;\;\;\; & 0\;\;\;\;\;\;\;\;\;\;\; & 0\\
   0\;\;\;\;\;\;\;\;\;\;\; & 0 \;\;\;\;\;\;\;\;\;\;\;&0 \;\;\;\;\;\;\;\;\;\;\; & 0\\
    0\;\;\;\;\;\;\;\;\;\;\; & 0 \;\;\;\;\;\;\;\;\;\;\;& 0\;\;\;\;\;\;\;\;\;\;\; & 0
  \end{smallmatrix}\\
  \begin{smallmatrix}
   0\;\;\;\;\;\;\;\;\;\;\; & 0 \;\;\;\;\;\;\;\;\;\;\;&0 \;\;\;\;\;\;\;\;\;\;\; & 0\\
    0\;\;\;\;\;\;\;\;\;\;\; & 0 \;\;\;\;\;\;\;\;\;\;\;& 0\;\;\;\;\;\;\;\;\;\;\; & 0
  \end{smallmatrix}&
  \begin{smallmatrix}
   0\;\;\;\;\;\;\;\;\;\;\; & 0 \;\;\;\;\;\;\;\;\;\;\;&0 \;\;\;\;\;\;\;\;\;\;\; & 0\\
    0\;\;\;\;\;\;\;\;\;\;\; & 0 \;\;\;\;\;\;\;\;\;\;\;& 0\;\;\;\;\;\;\;\;\;\;\; & 0
  \end{smallmatrix}&
  \begin{smallmatrix} 
  0 & -\gamma_0pe^{-i\phi} \\
    -\gamma_0pe^{i\phi} & 0 
  \end{smallmatrix}
  &\begin{smallmatrix}
   0\;\;\;\;\;\;\;\;\;\;\; & 0 \;\;\;\;\;\;\;\;\;\;\;&0 \;\;\;\;\;\;\;\;\;\;\; & 0\\
    0\;\;\;\;\;\;\;\;\;\;\; & 0 \;\;\;\;\;\;\;\;\;\;\;& 0\;\;\;\;\;\;\;\;\;\;\; & 0
  \end{smallmatrix}\\ 
  \begin{smallmatrix}
    0\;\;\;\;\;\;\;\;\;\;\; & 0\;\;\;\;\;\;\;\;\;\;\;& 0\;\;\;\;\;\;\;\;\;\;\;& 0\\
  0\;\;\;\;\;\;\;\;\;\;\; & 0\;\;\;\;\;\;\;\;\;\;\; & 0\;\;\;\;\;\;\;\;\;\;\; & 0\\
   0\;\;\;\;\;\;\;\;\;\;\; & 0 \;\;\;\;\;\;\;\;\;\;\;&0 \;\;\;\;\;\;\;\;\;\;\; & 0\\
    0\;\;\;\;\;\;\;\;\;\;\; & 0 \;\;\;\;\;\;\;\;\;\;\;& 0\;\;\;\;\;\;\;\;\;\;\; & 0
  \end{smallmatrix} & \begin{smallmatrix}
    0\;\;\;\;\;\;\;\;\;\;\; & 0\;\;\;\;\;\;\;\;\;\;\;& 0\;\;\;\;\;\;\;\;\;\;\;& 0\\
  0\;\;\;\;\;\;\;\;\;\;\; & 0\;\;\;\;\;\;\;\;\;\;\; & 0\;\;\;\;\;\;\;\;\;\;\; & 0\\
   0\;\;\;\;\;\;\;\;\;\;\; & 0 \;\;\;\;\;\;\;\;\;\;\;&0 \;\;\;\;\;\;\;\;\;\;\; & 0\\
    0\;\;\;\;\;\;\;\;\;\;\; & 0 \;\;\;\;\;\;\;\;\;\;\;& 0\;\;\;\;\;\;\;\;\;\;\; & 0
  \end{smallmatrix} & \begin{smallmatrix}
  0\;\;\;\;\;\;\;\;\;\;\; &  0\\
  0\;\;\;\;\;\;\;\;\;\;\; &  0\\
   0\;\;\;\;\;\;\;\;\;\;\; &  0\\
    0\;\;\;\;\;\;\;\;\;\;\; &   0
   \end{smallmatrix}  & 
  \begin{smallmatrix}
  \varepsilon_{2,eff}&-\gamma_0pe^{-i\phi}&0& \gamma_{2,eff}\\
   -\gamma_0pe^{i\phi}&0&0&0\\
 0&0&0&-\gamma_0pe^{-i\phi}\\
 \gamma_{2,eff}&0&-\gamma_0pe^{i\phi}&\varepsilon_{2,eff}
  \end{smallmatrix}
  \end{smallmatrix}
\right)  
\end{align}
\end{landscape}

{\bf Acknowledgements}\\
 
This work is supported in part by University of Lodz.

\end{document}